\documentclass[12pt,preprint]{aastex}

\newcounter{thefigs}

\newcounter{thetabs}

\newcounter{address}

\def\simless{\mathbin{\lower 3pt\hbox
	{$\,\rlap{\raise 5pt\hbox{$\char'074$}}\mathchar"7218\,$}}} 
\def\simgreat{\mathbin{\lower 3pt\hbox
	{$\,\rlap{\raise 5pt\hbox{$\char'076$}}\mathchar"7218\,$}}} 

\slugcomment{Ready to submit \aj}

\begin{document}
 

\title{The Cut \& Enhance method : selecting clusters of
galaxies from the SDSS commissioning data.
}

\author{
Tomotsugu Goto\altaffilmark{\ref{CosmicRay}}\altaffilmark{\ref{CarnegieMellon}}, 
Maki Sekiguchi\altaffilmark{\ref{CosmicRay}}, 
Robert C. Nichol\altaffilmark{\ref{CarnegieMellon}},
Neta A. Bahcall\altaffilmark{\ref{Princeton}},
Rita S.J. Kim\altaffilmark{\ref{Princeton}},
James Annis\altaffilmark{\ref{Fermilab}},
\v{Z}eljko Ivezi\'{c}\altaffilmark{\ref{Princeton}},
J. Brinkmann\altaffilmark{\ref{APO}},
Gregory S. Hennessy\altaffilmark{\ref{USNO}},
Gyula P.~Szokoly\altaffilmark{\ref{Potsdam}},
Douglas L. Tucker\altaffilmark{\ref{Fermilab}},
}
\setcounter{address}{1}
\altaffiltext{\theaddress}{
\stepcounter{address}
Institute for Cosmic Ray Research, University of Tokyo, Kashiwanoha, Kashiwa, Chiba 277-0882, Japan
\label{CosmicRay}}
\altaffiltext{\theaddress}{
\stepcounter{address}
Department of Physics, Carnegie Mellon University, 
5000 Forbes Avenue, Pittsburgh, PA 15213-3890 
\label{CarnegieMellon}} 
\altaffiltext{\theaddress}{
\stepcounter{address}
Princeton University Observatory, Princeton,
NJ 08544
\label{Princeton}}
\altaffiltext{\theaddress}{
\stepcounter{address}
Fermi National Accelerator Laboratory, P.O. Box 500,
Batavia, IL 60510
\label{Fermilab}}
\altaffiltext{\theaddress}{
\stepcounter{address}
Apache Point Observatory,
2001 Apache Point Road,
P.O. Box 59, Sunspot, NM 88349-0059
\label{APO}}
\altaffiltext{\theaddress}{
\stepcounter{address}
U.S. Naval Observatory,
3450 Massachusetts Ave., NW,
Washington, DC  20392-5420
\label{USNO}}
\altaffiltext{\theaddress}{
\stepcounter{address}
Astrophysikalisches Institut Potsdam,
An der Sternwarte 16, D-14482 Potsdam, Germany
\label{Potsdam}}
%


\begin{abstract}
	
 We describe an automated method, the Cut \& Enhance method (CE) 
for detecting clusters of galaxies in
 multi-color optical imaging surveys.
This method uses simple color cuts, combined with a
density enhancement algorithm, to up--weight pairs of galaxies that are close
in both angular separation and color. The method is semi--parametric since it
uses minimal assumptions about cluster properties in order to minimize
possible biases.  No assumptions are made about the shape of clusters, their
radial profile or their luminosity function.
 The method is successful in finding systems ranging from poor to rich clusters of galaxies, of both regular and irregular shape.
 We determine the selection function of the CE
method via extensive Monte Carlo simulations which use both the real, observed
background of galaxies and a randomized background of galaxies.  We use
 position shuffled and color shuffled data to perform the false positive test.
 We have also
 visually checked all the clusters detected by the CE method.

 We apply the CE method to the 350 deg$^2$ of the SDSS (Sloan
 Digital Sky Survey) commissioning data and construct a SDSS CE
 galaxy cluster catalog with an estimated redshift
 and richness for each cluster. 
 The CE method is compared with other
   cluster selection methods used on SDSS data such as the Matched Filter (Postman et al.\ 1996, Kim et al.\ 2001) , maxBCG  technique (Annis et al.\
    2001) and Voronoi Tessellation (Kim et al.\ 2001).  
 The CE method can be adopted for cluster selection in any multi-color
 imaging surveys.

\end{abstract}


\section{Introduction}
\label{Jun 28 09:03:18 2001}

 Clusters of galaxies are 
 the most massive virialized systems known and
 provide powerful tools in the study of cosmology and extragalactic
   astronomy.
 For
example, clusters are efficient tracers of the large--scale structure in the
Universe as well as determining the amount of dark matter on Mpc scales
 (Bahcall 1998; Carlberg et al. 1996; Borgani \& Guzzo 2000 and Nichol 2001 and references
therein).
 Furthermore, clusters provide a laboratory within which to study a
large number of galaxies at the same redshift and thus assess the effects of
dense environments on galaxy evolution {\it e.g.}  morphology--density
relation (Dressler et al.\ 1980, 1984, 1997), Butcher--Oemler effect (Butcher
\& Oemler 1978, 1984) and the density dependence of the luminosity function of
galaxies (Garilli et al. 1999).
 In recent years, surveys of clusters of galaxies have been used extensively in
constraining cosmological parameters such as $\Omega_m$, the mass density
parameter of the universe, and $\sigma_8$, the amplitude of mass fluctuations
at a scale of 8 $h^{-1}$ Mpc (see
 Oukbir \& Blanchard 1992; Viana \& Liddle
1996, 1999; Eke et al. 1996; Bahcall, Fan \& Cen 1997; Henry 1997, 2000;
Reichart et al. 1999 as examples of an extensive
literature on this subject).  Such constraints are achieved through the
comparison of the evolution of the mass function of galaxy clusters, as
predicted by the Press-Schechter formalism (see Jenkins et al. 2001 for the
latest analytical predictions) or simulations ({\it e.g.} Evrard et
 al. 2001 and Bode et al. 2001),
with the observed abundance of clusters with redshift. Therefore, to obtain
robust constraints on $\Omega_m$ and $\sigma_8$, we need large samples of
clusters that span a large range in redshift and mass as well as possessing a
well--determined selection function (see Nichol 2001).

 Despite their importance, existing catalogs of clusters are limited in both
their size and quality. For example, the Abell catalog of rich clusters (Abell
1958), and its southern extension (Abell, Corwin and Olowin 1989), are still
some of the most commonly used catalogs in astronomical research even though
they were constructed by visual inspection of photographic plates.  Another
large cluster catalog by Zwicky et al.\ (1961-1968) was similarly constructed
by visual inspection.  Although the human eye can be efficient in detecting
galaxy clusters, it suffers from subjectivity and incompleteness. 
 For cosmological studies, the major disadvantage of visually
constructed catalogs is the difficulty to quantify selection bias
and thus, the selection function. 
 Furthermore, the response of photographic
 plates is not uniform. 
 Plate-to-plate sensitivity variations can
 disturb the uniformity of the
 catalog.
 To overcome these problems,
 several cluster catalogs have been constructed using
 automated
 detection methods on CCD imaging data. 
 They
 have been, however,  restricted to small areas
 due to the lack of large--format CCDs.
 {\it e.g.} the PDCS catalog (Postman et al.\ 1996) only covers 5.1 deg$^2$
 with 79 galaxy clusters. 
 The need for a uniform, large cluster catalog is strong.
 The Sloan Digital Sky Survey (SDSS; York et al.\ 2000)
 data offer the opportunity 
 to produce the largest and most uniform galaxy cluster catalog in
 existence
 because
 the SDSS
 is the largest CCD imaging
  survey currently underway
  scanning 10,000 deg$^2$ centered approximately on the North Galactic
 Pole. 

 The quantity and quality of the SDSS data demands the use of sophisticated
cluster finding algorithms to help maximize the number of true cluster
detections while suppressing the number of false positives.
 The history of the automated cluster finding methods goes back to Shectman's
count-in-cell method (1985). He counted the number of galaxies in cells
 on the sky to estimate the galaxy density. Although this provided important
progress over the visual inspection, the results depend on the size
and position of the cell.
 Currently the commonly used automated cluster finding method is the Matched
Filter technique (MF) (Postman et al.\ 1996, Kawasaki et 
al.\ 1998,  Kepner et al.\ 1999, Schuecker \& Bohringer 1998, Bramel et
al.\ 2000, Lobo et al.\ 2000, da Costa et al.\ 2000 and Willick et al.\ 2000).
 The method assumes a filter
 for the 
 radial profile of galaxy clusters and for the luminosity function of
 their members.
 It then selects clusters from imaging data by maximizing the likelihood
 of matching the data to the cluster model.
Although the method has been successful, galaxy clusters that do
not fit the model assumption (density profile and LF) may be missed.
  We present here a
new cluster finding method called the {\it Cut } and {\it Enhancement }
method, or CE.
 This new algorithm is semi--parametric and is designed to be as
simple as possible using the minimum number of assumptions possible about
cluster properties. In this way, it should be sensitive to all types of galaxy
overdensities even those that may have recent under--gone a merger and
therefore, are highly non--spherical.
 One major difference between CE and previous cluster finders is that CE
 makes full use of colors of galaxies, which become available due to the
 advent of the accurate CCD photometry of the SDSS data.  
 We apply this detection method 
 on 350 deg$^2$ of the SDSS commissioning
 data and  construct the large cluster catalog.
 The catalog ranges from rich clusters to the more numerous poor clusters
 of galaxies over this area.  We also determine the selection function of the CE method.

In Sect.\ \ref{sec:data}, we describe the SDSS commissioning data. 
In Sect.\ \ref{sec:method}, we describe the detection strategy of Cut \& Enhance method. 
In Sect.\ \ref{sec:monte}, we present the performance test of the Cut \& Enhance method and selection
function using Monte Carlo simulations.
In Sect.\ \ref{sec:visual}, we visually check the success rate of the Cut \& Enhance method.
In Sect.\ \ref{sec:compare}, we compare Cut \& Enhance method with the
 other detection methods applied to the SDSS
data. 
In Sect.\ \ref{sec:summary}, we summarize the results.


\section{The SDSS commissioning data }\label{sec:data}

 The data we use to construct the SDSS Cut \& Enhance galaxy cluster catalog 
are equatorial scan data taken in September 1998 and
March 1999 during the early part of the SDSS commissioning phase.
A contiguous area of 250 deg$^2$ (145.1$<$RA$<$236.0, -1.25$<$DEC$<$+1.25) and 150 deg$^2$ (350.5$<$RA$<$56.51, -1.25$<$DEC$<$+1.25)  were obtained during four nights,
where seeing varied from 1.1'' to 2.5''. 
 Since we intend to use the CE method at the faint end of imaging data,
 we include galaxies to
$r^*$=21.5 (petrosian magnitude), which is the star/galaxy separation
limit.  
 Since the SDSS photometric system is not yet finalized, we refer to the
 SDSS photometry presented here as $u^*,g^*,r^*,i^*$ and $z^*$.
 The technical aspects of the SDSS camera are
 described in Gunn et al.\ (1998).  Fukugita et al.\ (1996) describe the
 color filters of SDSS. The details about SDSS commissioning data are
 described in Stoughton et al. (2001).

\section{Cut \& Enhance cluster detection method}\label{sec:method}

\subsection{Color cut}

 The aim of the Cut \& Enhance 
 method is to construct a
 cluster catalog that has little bias as possible by minimizing the
 assumptions about cluster properties.
 If a method assumes
 a luminosity function or radial profile, for example, the resulting clusters will
be biased to the detection model used.
 We thus exclude all such assumptions except for a generous color cut.
 The assumption on colors of cluster galaxies appears to be robust,
  as all the
galaxy clusters appear to have the same general color-magnitude relation
(Gladders et al.\ 2000). 
 Even a claimed ``dark cluster''
(Hattori et al.\ 1997) was found to have a normal color magnitude relation.
(Benitez et al.\ 1999, Clowe et al.\ 2000, Soucail et al.\ 2000).
 
\begin{figure}
\plotone{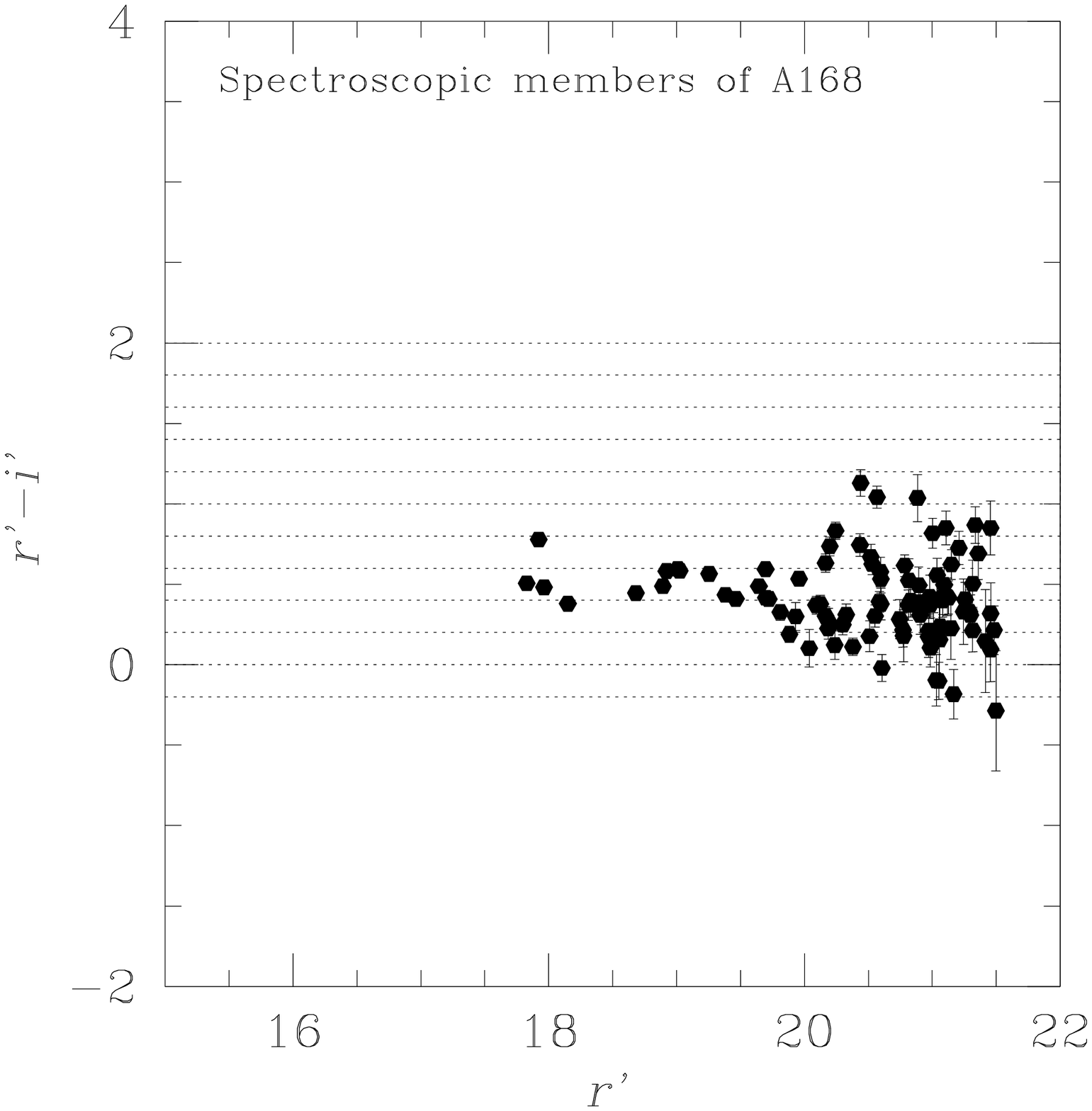}
\caption{
\label{fig:grcut168}
$r^*-i^*$ color-magnitude diagram of A168.
$r^*-i^*$ color is plotted against $r^*$ magnitude for confirmed member
 galaxies of A168.
Colors and magnitude are taken from the SDSS commissioning data by matching
 up the positions with  the spectroscopic observation of Katgert et al.\ (1998).
The standard errors of colors estimated by the reduction
 software are shown as error bars.
$r^*-i^*$ color cut bins are superimposed on the color-magnitude relation of Abell 168.
Horizontal dotted lines are the borders of the
 color cuts.
}
\end{figure}

\begin{figure}
\plotone{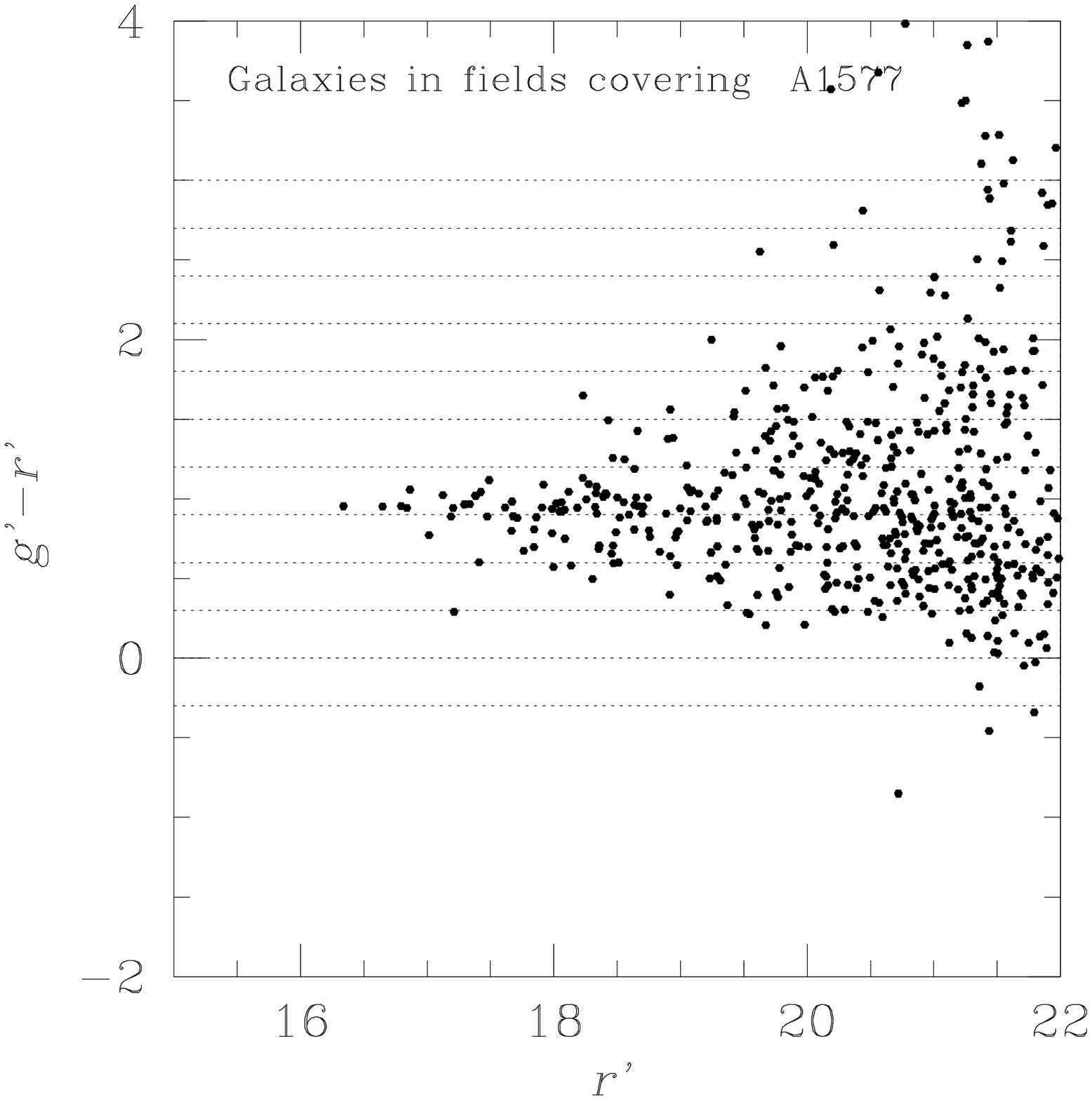}
\caption{
\label{fig:grcut}
$g^*-r^*$ color-magnitude diagram.
$g^*-r^*$ color cut bins are
superimposed on the color-magnitude relation of Abell 1577.
 The abscissa is the $r^*$ apparent magnitude. The ordinate is $g^*-r^*$ color. 
Galaxies in the SDSS fields covering A1577 ($\sim8.3\times10^{-2}$ deg$^2$) are plotted. 
Data are taken from the SDSS commissioning data.
Horizontal dashed lines are the borders of the
 color cuts.
}
\end{figure}

\begin{figure}
\plotone{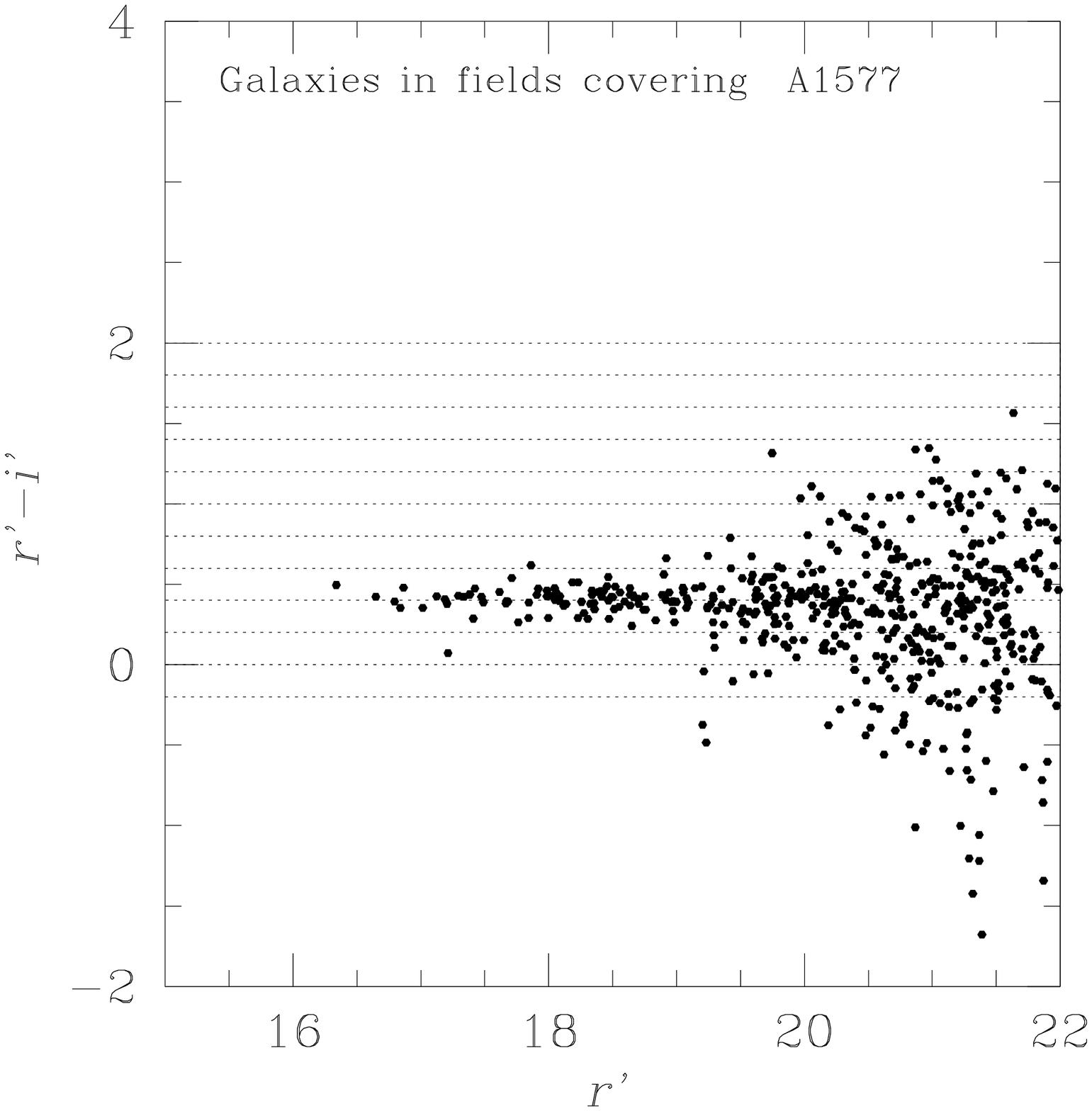}
\caption{
\label{fig:ricut}
$r^*-i^*$ color-magnitude diagram.
$r^*-i^*$ color cut bins are
superimposed on the color-magnitude relation of Abell 1577.
 The abscissa is the $r^*$ apparent magnitude. The ordinate is $r^*-i^*$ color. 
Galaxies in the SDSS fields  covering A1577 ($\sim8.3\times10^{-2}$ deg$^2$) are plotted. 
Colors and magnitudes are taken from the SDSS commissioning data.
Horizontal dashed lines are the borders of the
 color cuts.
}
\end{figure}

\begin{figure}
\plotone{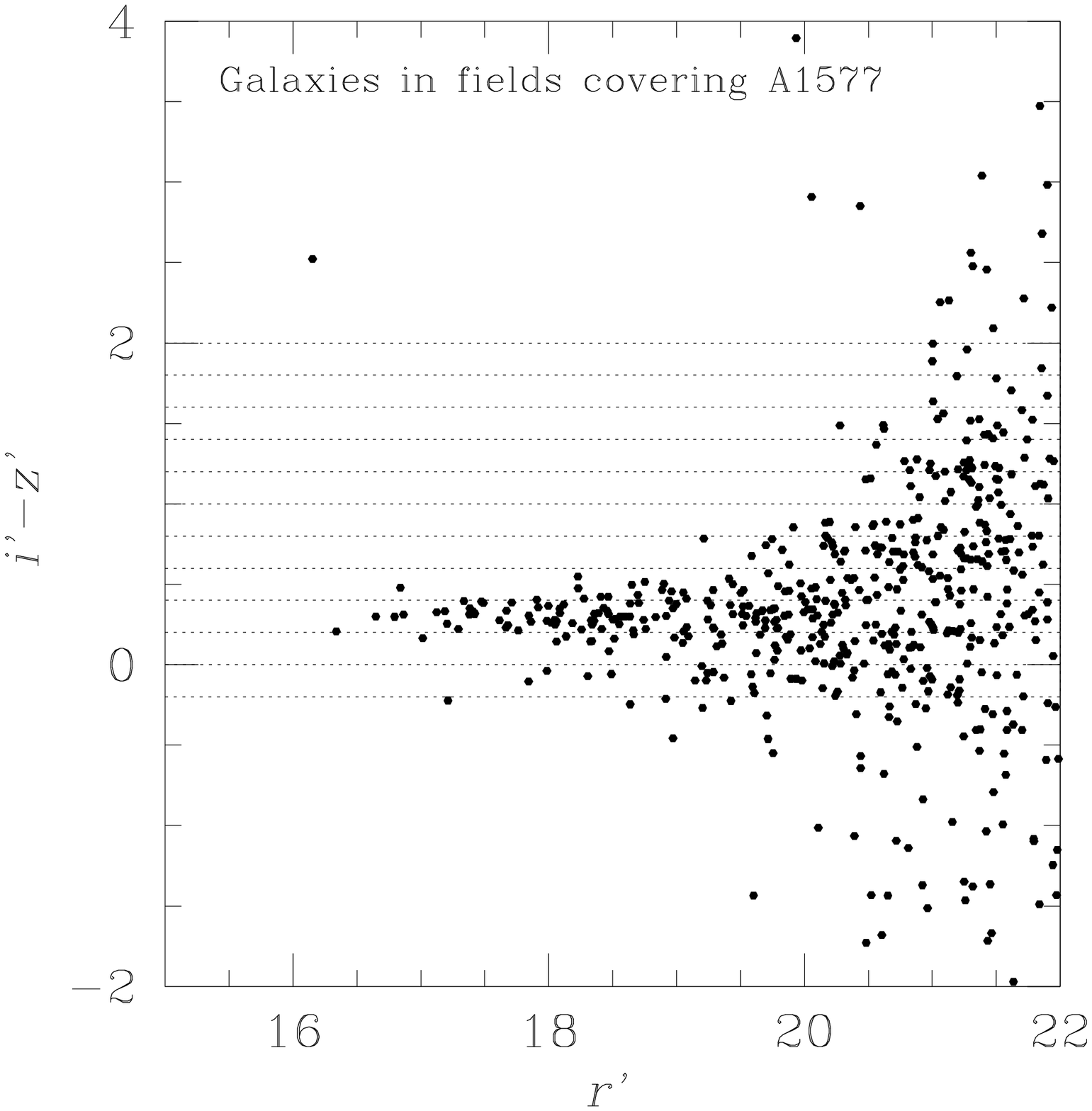}
\caption{
\label{fig:izcut}
$i^*-z^*$ color-magnitude diagram.
$i^*-z^*$ color cut bins are
superimposed on the color-magnitude relation of Abell 1577.
 The abscissa is the $r^*$ apparent magnitude. The ordinate is $i^*-z^*$ color. 
Galaxies in the SDSS fields  covering A1577 ($\sim8.3\times10^{-2}$ deg$^2$) are plotted. 
Colors and magnitudes are taken from the SDSS commissioning data.
Horizontal dashed lines are the borders of the
 color cuts.
}
\end{figure}

 Galaxy clusters are known to have a tight color-magnitude
relation; among the various galaxy populations within a cluster,
(i.e. spiral, elliptical, dwarf, irregular),
 bright red elliptical galaxies have similar color and 
they populate a red ridge line 
 in the color-magnitude diagram (called the color-magnitude relation).
Bower, Lucey, \& Ellis (1992) obtained high precision $U$ and $V$ photometry
of spheroidal galaxies in two local clusters, Virgo and Coma. They
observed a very small scatter, $\delta(U-V)<0.035$ rms. 
Ellis et al.\ (1997)
studied the $U-V$ color-magnitude relation at high redshift ($z\sim$0.54) and found a
scatter of $<$ 0.1 mag rms. Similarly,  Stanford et al.\ (1998) studied optical-infrared
colors ($R-K$) of early-type (E+S0) galaxies in 19 galaxy clusters out to $z$=0.9
and found a very small dispersion in the optical-infrared 
colors of $\sim$0.1 mag rms.
Fig.\ \ref{fig:grcut168} shows the color-magnitude diagram in ($r^*-i^*$
 vs. $r^*$) using SDSS data for galaxy members in the cluster A168 ($z$=0.044).
The member galaxies are identified by matching 
 the positions of galaxies in the SDSS commissioning data with the
 spectroscopic observation of 
Katgert et al.\ (1998). 
The error bars show the standard errors of $r^*-i^*$ color estimated by
the SDSS reduction
 software (Lupton et al.\ 2001). 
The red ridge line of the color-magnitude relation
is seen at $r^*-i^*\sim$0.4 from $r^*=17.5$ to $r^*=20$.
The scatter is 0.08 mag from the brightest to $r^*=18$.
Fig.\ \ref{fig:grcut} shows the color-magnitude diagram (in $g^*-r^*$ vs
$r^*$) for all galaxies in the SDSS fields ($\sim8.3\times10^{-2}$ deg$^2$) 
that contain Abell 1577. A1577 has a redshift of $z\sim0.14$ and Abell richness class $\sim1$.
Fig.\ \ref{fig:ricut} and Fig.\ \ref{fig:izcut} show color-magnitude
 diagrams for the same field in $r^*-i^*$ and $i^*-z^*$ colors, respectively.
 All the galaxies in the region are included.
Even without spectroscopic information,
 the red ridge of the color-magnitude relation is clearly visible
 as the  horizontal distribution of galaxy colors.
The scatter in the color-magnitude relation is the largest in $g^*-r^*$
  because the difference of the galaxy SED due to  the age or
  metallicity difference is prominent around
  $3500\sim5000$\AA.
The color distribution is much wider at faint
 magnitudes, partly because fainter galaxies have larger color errors,
and partly because of the increase in the number of background galaxies.

The color-magnitude relation is known to have a slight tilt (Kodama et
 al.\ 1998).
 The tilt is small in the SDSS color bands.
 The tilt and its scatter in the case of A1577 (Fig.\ \ref{fig:grcut})
 is summarized in Table \ref{tab:cm-tilt}.
 The tilt is small in $g^*-r^*$ and $r^*-i^*$ ($\sim$0.08), and even smaller 
 in $i^*-z^*$ (0.0018). These values are much smaller than the color cuts
 of CE. The scatters are also small: 0.081, 0.040 and 0.033 in $g^*-r^*$, $r^*-i^*$ and
 $i^*-z^*$, well smaller than the color cuts of CE. 
 The small scatter of $<$0.1 mag is consistent with the previous 
works (Bower et al.\ 1992; Ellis et al.\ 1997; Stanford et al.\ 1998).
The tilt of the color-magnitude relation is smaller than the scatter 
in the SDSS color bands. 

%
%

\begin{figure}
\begin{center}
\plotone{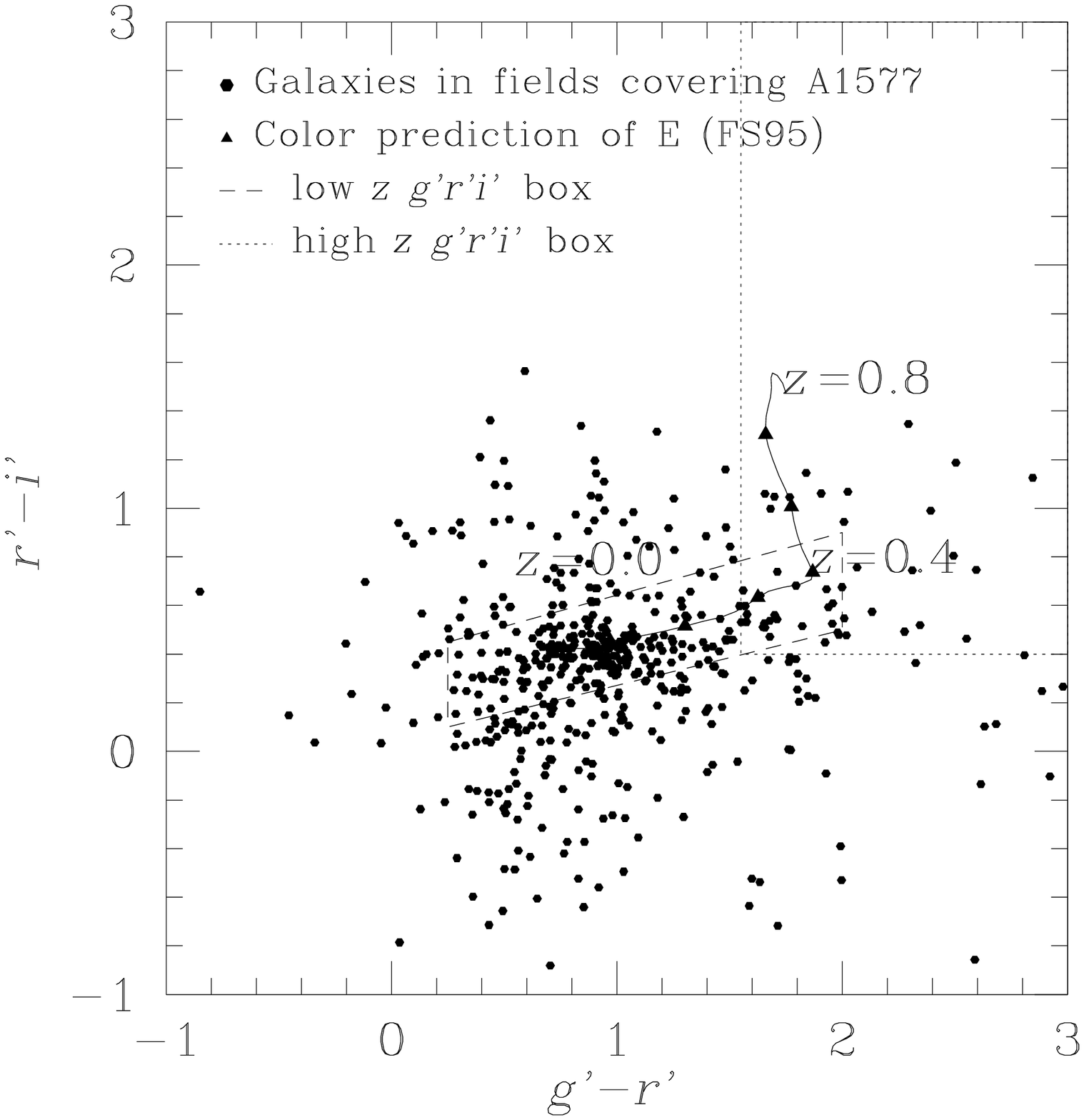}
 \caption{
 \label{fig:griboxes1}
 $g^*-r^*-i^*$ color-color boxes to find galaxy clusters.
 The abscissa is the $g^*-r^*$ color. 
 The ordinate is $r^*-i^*$ color.
The low-$z$ $g^*-r^*-i^*$ box is  drawn with dashed lines.
The high-$z$ $g^*-r^*-i^*$ box is  drawn with dotted lines. 
  Galaxies brighter than $r^*=22$ in the SDSS fields
 ($\sim8.3\times10^{-2}$deg$^2$)  which covers A1577 are plotted with
 small dots. 
 The triangles show the 
 color prediction of elliptical galaxies (Fukugita et al.\ 1995).
 }
\end{center}
\end{figure}

 We use the color-magnitude relation to enhance the detection signal of
 galaxy clusters. Such a use of colors of galaxies become possible only
 recently due to the appearance of large CCD based data ($e.g.$ SDSS). 
 Since cluster members have similar colors, 
 we use specific but generous color cuts, 
 to enhance the contrast of galaxy clusters.
The colors of 
red elliptical galaxies change as a function of redshift.
Fig.\ \ref{fig:griboxes1} presents the color-color diagram,
 $g^*-r^*$ vs $r^*-i^*$, for all 
galaxies brighter than $r^*$=22 in the SDSS fields
that covers A1577,
 as well as the color predictions of elliptical galaxies at different
 redshifts (shown by the triangles; Fukugita et al.\ 1995).   
 The $g^*-r^*$ color becomes redder from $z$=0 to $z$=0.4 and 
 $r^*-i^*$ reddens monotonously. 
At $z\sim0.4$, the 4000 $\AA$  
 break of an elliptical galaxy
 crosses the border between $g^*$ and $r^*$ bands, and 
appears as a sharp turn in the color at this redshift (Fig.\
 \ref{fig:griboxes1}). By using this color change, we can reject
 foreground and background galaxies and can select galaxies likely to be
 in a certain redshift range in the following way. This is a big
 advantage of having multi-color data since optical cluster finders have
 suffered chance projections of galaxies in the sky.
To select galaxies with similar colors,
we divide the $g^*-r^*$ vs. $r^*$ color-magnitude diagram into eleven bins. 
The bins are shown in Fig.\ \ref{fig:grcut} as horizontal dashed lines.
The bins are not tilted because the tilt is almost negligible in the SDSS bands (see above), and 
 because we wish to minimize the assumptions used for cluster selection.
 Any specific color bin reflects the redshifts of the cluster:
 Blue color bins represent low redshift clusters while red bins represent 
 higher redshift clusters.
 We use two bins as one color cut in order to produce overlap in the
 color cuts;
 the cut is shifted by one bin each time we step to a higher redshift
(redder cut).

 Similarly, we use ten color cuts in both $r^*-i^*$ (shown in Fig.\ \ref{fig:ricut}
 as dashed lines) and ten color
 cuts in $i^*-z^*$ (shown in Fig.\ \ref{fig:izcut} as dashed lines).
The width of the bins in $g^*-r^*$, $r^*-i^*$ and $i^*-z^*$ color are 0.2 mag,
 0.1 mag and 0.1 mag, respectively.
The width of the $r^*-i^*$ and $i^*-z^*$ bins is smaller than the $g^*-r^*$
 width
because the colors of elliptical galaxies have less scatter in $r^*-i^*$ and
 $i^*-z^*$ than in $g^*-r^*$. 
The above color cuts, in the three colors, are applied independently.
Galaxies which have color errors larger than the size of
 the color bin are rejected. 
The standard color error estimated by the SDSS reduction software
 at $r^*=21.5$  (the limiting magnitude used in the Cut \& Enhance method.)
 is  0.20$\pm$0.09, 0.16$\pm$0.06 and 0.26$\pm$0.1 in $g^*-r^*$, $r^*-i^*$
 and $i^*-z^*$, respectively. In $g^*-r^*$ and $r^*-i^*$, the color error
  is smaller
 than the size of the color cut box. In $i^*-z^*$, the color error at
 $r^*=21.5$ is
 slightly larger than the size of the color cut boxes (0.2 mag), at $r^*$=20.5
, however, 
the errors of $i^*-z^*$ is 0.11$\pm$0.05.

\begin{figure}
\begin{center}
\plotone{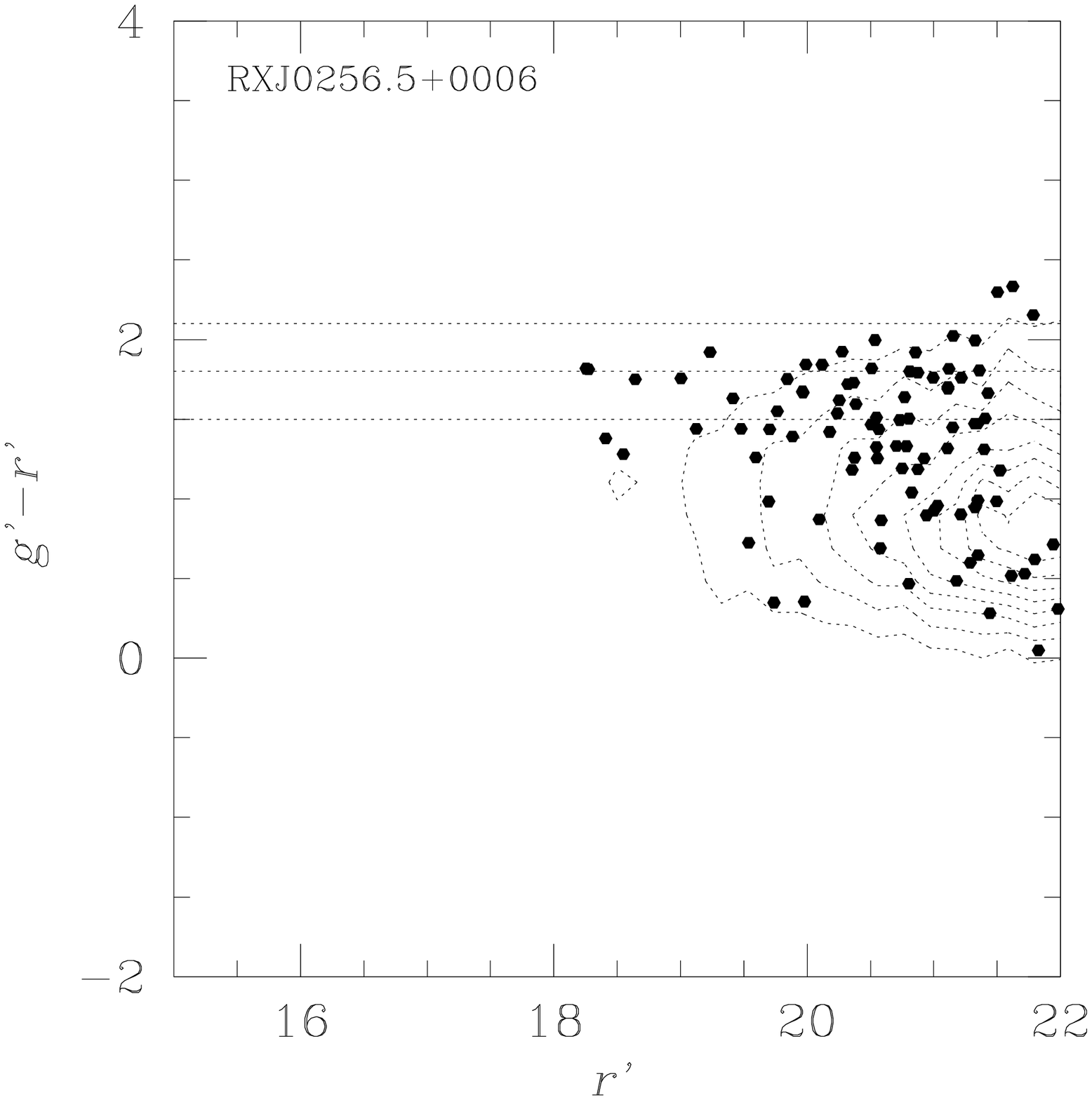}
 \caption{
 \label{fig:RXJ0256_gr}
 An example of color-cut capturing color-magnitude relation.
 Galaxies within 1.5$h^{-1}$Mpc aparture around RXJ0256.5+0006
 ($z$=0.36) are plotted as black dots.
 Distribution of all the galaxies in the SDSS commissioning data is
 drawn as contours.
 The $g^*-r^*$ color-cut successfully caputuring red-sequence of RXJ0256.5+0006
 and removing the foreground galaxies bluer than the sequence.
 }
\end{center}
\end{figure}

\begin{figure}
\begin{center}
\plotone{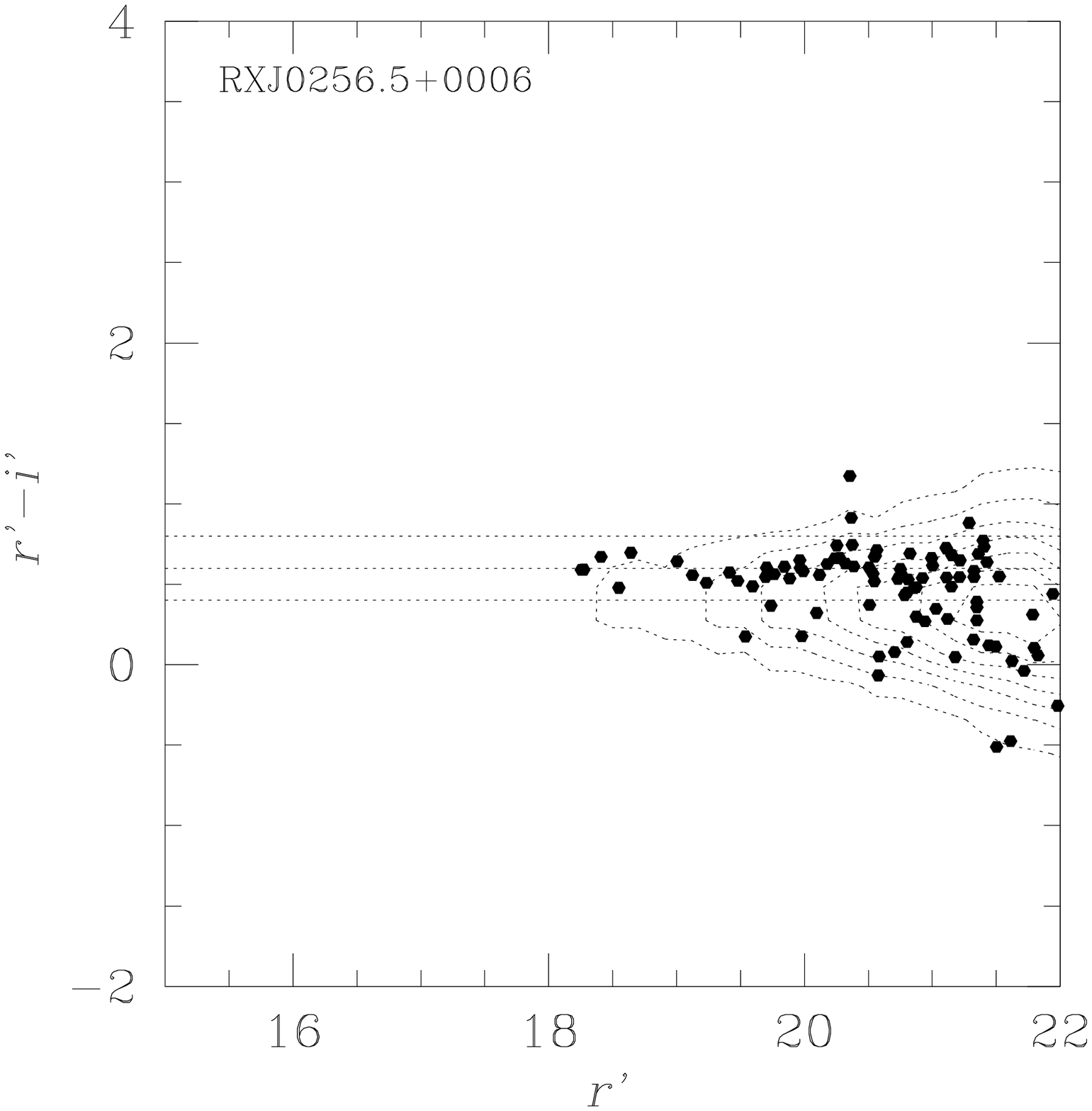}
 \caption{
 \label{fig:RXJ0256_ri}
 An example of color-cut capturing color-magnitude relation.
 Galaxies within 1.5$h^{-1}$Mpc aparture around RXJ0256.5+0006
 ($z$=0.36) are plotted as black dots.
 Distribution of all the galaxies in the SDSS commissioning data is
 drawn as contours.
 The $r^*-i^*$ color-cut successfully caputuring red-sequence of RXJ0256.5+0006
 and removing the foreground galaxies bluer than the sequence.
 }
\end{center}
\end{figure}

 In Fig. \ref{fig:RXJ0256_gr} and Fig. \ref{fig:RXJ0256_ri}, we
 demonstrate the effect of the color cut. Black dots are the galaxies
 within 2.7' (1.5$h^{-1}$Mpc at $z$=0.37) from the center of
 RXJ0256.5+0006 (Romer et al. 2001) . No background and foreground correction are applied. 
 Contours represent the distribution of all the galaxies of the SDSS
 imaging data. The corresponding color cuts to the redshift of the cluster are drawn in each figure. 
 In each case, the color cuts capture the red-sequence of RXJ0256.5+0006 successfully
 and reject foreground galaxies as designed. In fact, we show
in Table
 \ref{tab:RXJ0256_color_cut}, 
the  fraction of galaxies inside of the color
 cut for both in cluster region and outside of cluster region.
 As shown in the Fig. \ref{fig:RXJ0256_gr} and
 Fig. \ref{fig:RXJ0256_ri}, indeed the fraction in the color cut
 increases dramatically from 13.5\% to 36.9\% in $g^*-r^*$ cut and from 42.4\% to
 62.1\% in $r^*-i^*$ cut. The efficiency of color cut increases as we see
 higher redshift apart from the foreground color distribution of
 galaxies.
 The upper left panel in Fig. \ref{fig:beforecut_RXJ} shows  the galaxy distribution of the SDSS commissioning data around RXJ0256.5+0006
 before applying any cut. The upper right panel shows the galaxy distribution after applying the $g^*-r^*$ color cut at the cluster redshift, it illustrates the color cut enhancement of  the cluster.

\subsection{Color-color cut}

\begin{figure}
\begin{center}
\plotone{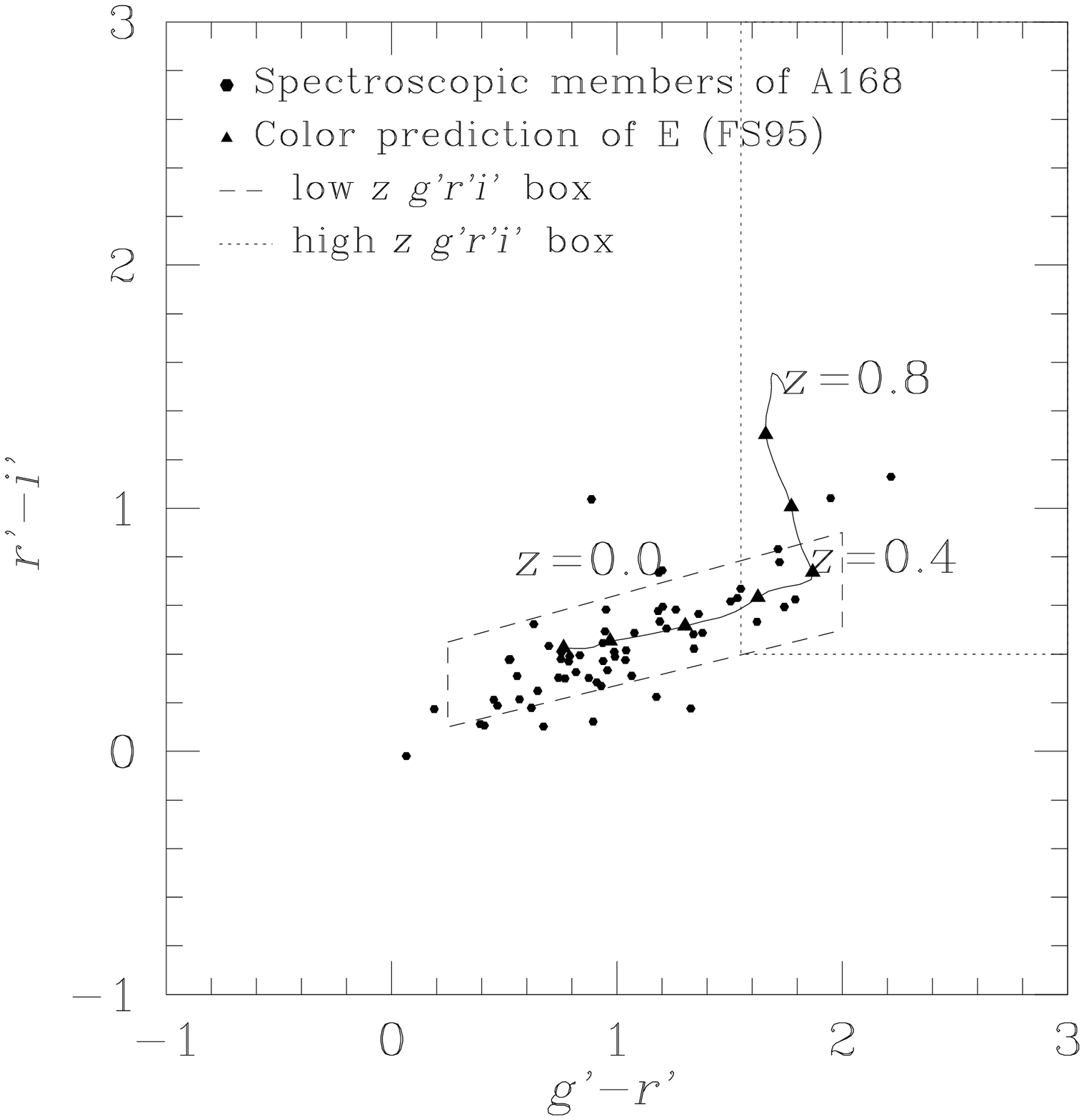}
 \caption{
 \label{fig:griboxes_a168}
Color-color diagram of spectroscopically confirmed member galaxies of
 A168.
 The abscissa is $g^*-r^*$ color. 
 The ordinate is $r^*-i^*$ color.
The low-$z$ $g^*-r^*-i^*$ box is  drawn with dashed lines.
The high-$z$ $g^*-r^*-i^*$ box is  drawn with dotted lines. 
 Galaxies brighter than $r^*=$21 which matched up the spectroscopically
 confirmed galaxies (Katgert et al.\ 1998) are plotted with dots. 
 The triangles show the 
 color prediction of elliptical galaxies (Fukugita et al.\ 1995).
}
\end{center}
\end{figure}

\begin{figure}
\begin{center}
\plotone{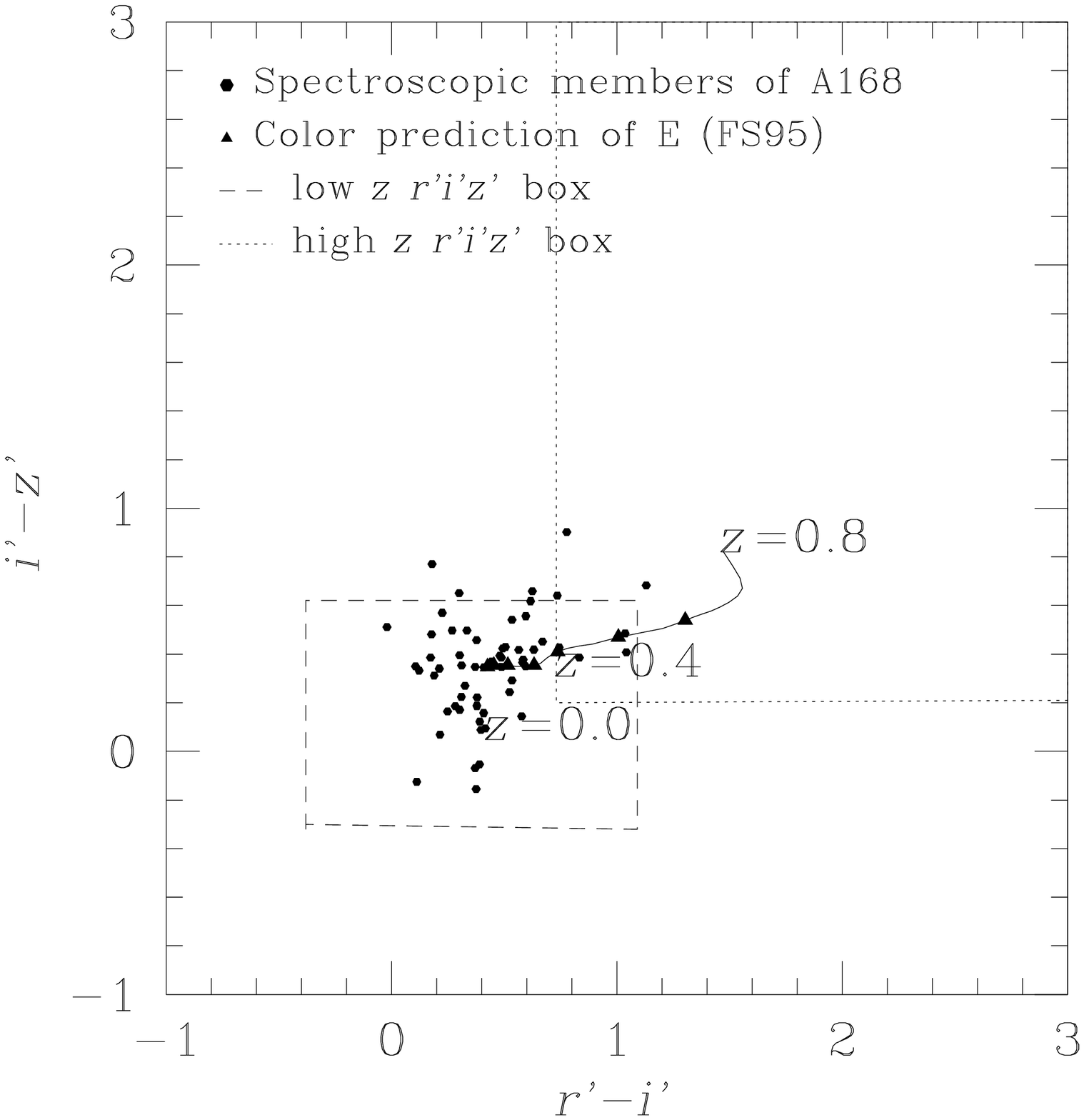}
	       \caption{
	       \label{fig:colors_of_Elliptical_a168}
Color-color diagram of spectroscopically confirmed member galaxies of
 A168.
 The abscissa is $r^*-i^*$ color. 
 The ordinate is $i^*-z^*$ color.
The low-$z$ $r^*-i^*-z^*$ box is  drawn with dashed lines.
The high-$z$ $r^*-i^*-z^*$ box is  drawn with dotted lines. 
 Galaxies brighter than $r^*=$21 which matched up the spectroscopically
 confirmed galaxies (Katgert et al.\ 1998) are plotted with dots. 
 The triangles show the 
 color prediction of elliptical galaxies (Fukugita et al.\ 1995).
}	      
\end{center}
\end{figure}

When more than two colors are available,
 it is more
effective to select galaxies in color-color space. 
 We thus added four additional color-color-cut boxes to enhance the contrast
 of galaxy clusters.
The cuts are low-$z$ and high-$z$  boxes in
$g^*-r^*-i^*$ space 
and in $r^*-i^*-z^*$ space, as shown in Fig.\ \ref{fig:griboxes_a168} and Fig.\ \ref{fig:colors_of_Elliptical_a168}.
 These color boxes are based on the fact that cluster
 galaxies concentrate 
 in specific regions in
 color-color space (Dressler \& Gunn 1992).
 In Fig.\ \ref{fig:griboxes_a168}, we show the $g^*-r^*$ vs. $r^*-i^*$ 
  color-color diagram of A168 for
 spectroscopically confirmed member galaxies (Katgert et al.\ 1998) brighter than $r^*$=21. 
 The low-$z$ $g^*-r^*-i^*$ color-color-cut box is shown with dashed lines and
 the high-$z$ $g^*-r^*-i^*$ color-color-cut box is shown by the dotted
 lines.
 The triangles present the color prediction as a function of redshift
 for elliptical galaxies ($\Delta z$=0.1; Fukugita et al.\ 1995). 
 The scatter in the plots comes from the mixture from the
 different type of morphology.
 Similar results are shown in Fig.\ \ref{fig:colors_of_Elliptical_a168}
 for the $r^*-i^*-z^*$ color-color diagram of A168.
Member galaxies of A168 ($z$=0.044, Struble \& Rood 1999) are well
centered in the low-$z$ $g^*-r^*-i^*$
and $r^*-i^*-z^*$ boxes.

\begin{figure}
\begin{center}

\plotone{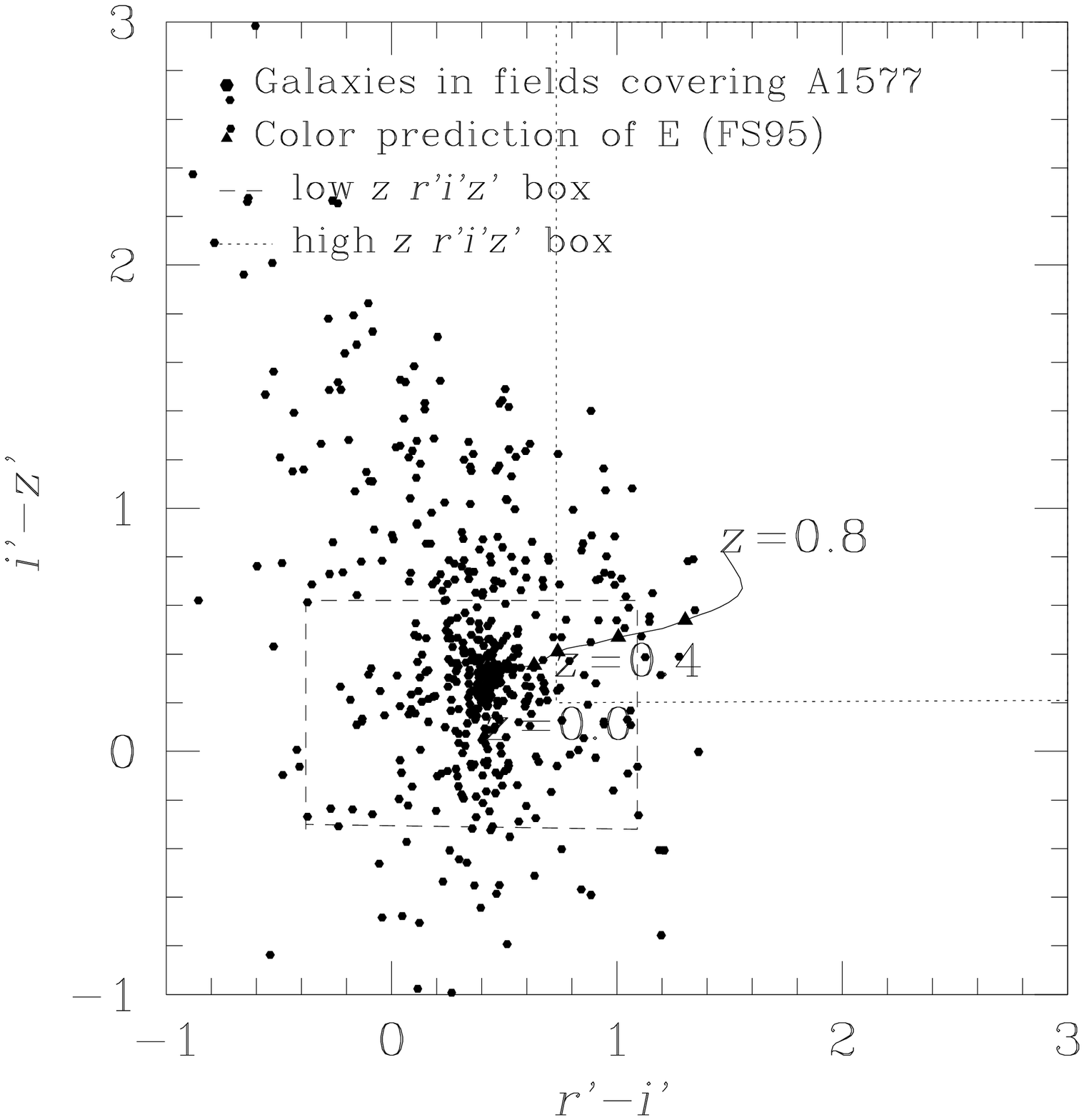}	
       \caption{\label{fig:colors_of_Elliptical}
 $r^*-i^*-z^*$ color-color boxes to find galaxy clusters.
 The abscissa is the $r^*-i^*$ color. 
 The ordinate is $i^*-z^*$ color.
The low-$z$ $r^*-i^*-z^*$ box is  drawn with dashed lines.
The high-$z$ $r^*-i^*-z^*$ box is  drawn with dotted lines. 
 Galaxies brighter than $r^*$=22 in the SDSS fields
 ($\sim8.3\times10^{-2} deg^2$)  which covers A1577 are plotted with small dots.
 The triangles show the 
 color prediction of elliptical galaxies (Fukugita et al.\ 1995).
}	      
\end{center}
\end{figure}

Fig.\ \ref{fig:griboxes1} is the $g^*-r^*-i^*$ color-color diagram of galaxies
 (brighter than $r^*$=22) in the 
 SDSS fields covering A1577 ($z$=0.14).
The low-$z$  and high-$z$ color-color-cut boxes are also
 shown.
The triangle points show the color prediction for 
 elliptical galaxies. 
Fig.\ \ref{fig:colors_of_Elliptical} represents similar results in the 
 $r^*-i^*-z^*$ color-color space for the same field.
Even though both cluster members and field galaxies are included in the plot,
the concentration of cluster galaxies inside the low-$z$ boxes is
clearly seen.

The color-color cuts are made based on the spectroscopic observation
 of Dressler \& Gunn (1992) and the color prediction of elliptical galaxies
 (Fukugita et al.\ 1995). We reject galaxies that have standard color
 errors larger than the size of the color-color boxes.
The standard color error at $r^*$=21.5 (the limiting magnitude of CE method.)
 is 0.20$\pm$0.09, 0.16$\pm$0.06 and 0.26$\pm$0.1 in $g^*-r^*$, $r^*-i^*$
 and $i^*-z^*$, respectively.
The smallest size of the color-color boxes is the $r^*-i^*$ side of the low-$z$
 $g^*-r^*-i^*$ box, which is 0.34 in $r^*-i^*$.
The standard color error is well within the color cut boxes even
 at $r^*$=21.5.
 In Fig. \ref{fig:beforecut_RXJ}, the upper left panel shows  the galaxy
 distribution of the SDSS commissioning data in 23.75 deg$^2$ before
 applying any cut. The upper right panel shows the galaxy distribution
 after applying the $g^*-r^*-i^*$ color-color cut.
 Abell clusters in the region are shown their position as numbers.
 It illustrates the color cut enhancement of nearby clusters. 
 We used RXJ0256.5+0006 ($z$=0.36) to numerate the fraction of inside of the color
 cut for both in cluster region 
 and outside of cluster region in Table
 \ref{tab:RXJ0256_color_cut}.  Indeed, 
 the fraction of galaxies in the color cut
 increases from 48.8\% to 58.3\% in $g^*-r^*-i^*$ cut and from 65.7\% to
 76.7\% in $r^*-i^*-z^*$ cut. Since the color cuts has overlaps at
 $z\sim$0.4, $g^*-r^*-i^*$ high$z$ cut also increases somewhat.




 We thus use 30 color cuts and four color-color cuts independently to
 search for clusters.
 We then merge 34 cluster candidate lists into a final cluster catalog.
Because of star/galaxy separation limit, we do not use galaxies fainter
than $r^*$=21.5 . 
The only main assumption made in the Cut \& Enhance detection method is
the above color cuts.

\begin{figure}
\begin{center}
\plotone{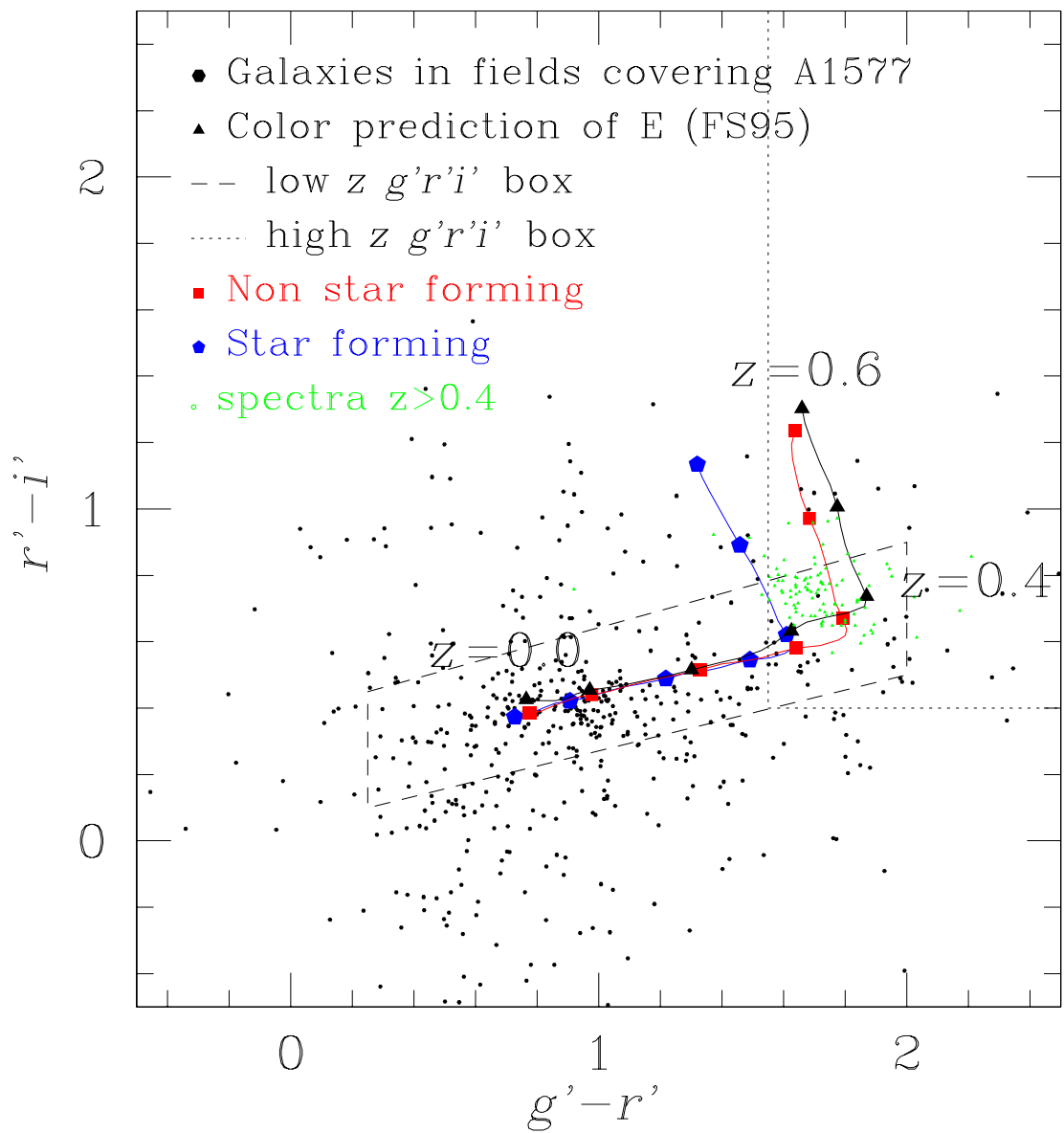}
\caption{
\label{fig:color-cut-test} 
Evaluation of high-z color cut. 
Filled triangles show the color prediction for Elliptical galaxies (Fukugita et al. 1995).
Open triangles show the color prediction of non star forming galaxies of
 PEGASE model (Fioc \& Rocca-Volmerange 1997).
Open squares show the color prediction of star forming galaxies of  PEGASE model.
Black dots are the galaxies around A1577.
High-$z$ color cut and low-$z$ color cut are drawn with dashed and
 long-dashed lines.
}
\end{center}
\end{figure}

 In Fig. \ref{fig:color-cut-test}, we plot the color prediction of galaxies
with evolving model with star formation (open triangle)  and the same model 
without star formation (open square) from $z$=0 to $z$=0.6 (PEGASE
model, Fioc, M., \& Rocca-Volmerange 1997). 
Filled triangles show the color prediction of elliptical galaxies
($\Delta z$=0.1;Fukugita et al. 1995).
The model galaxies with star formation are the extreme star forming galaxies. 
We plot
spectroscopic galaxies as gray dots. Black dots are the galaxies around
Abell 1577, for reference.
  Although the evolving model goes outside of the high-$z$ color cut 
 box at $z\sim$0.6, CE is designed to detect galaxy clusters if
 enough red galaxies (shown as triangles) are in the color cut by weighting the galaxies with
 similar color. 
 In fact, spectroscopic galaxies (shown as green dots, 0.4$<z\leq$0.5 )
 are well within the high-$z$ box (100 galaxies with $z>$0.4 are randomly taken from the
 SDSS spectroscopic data).
 As seen in the real catalog in Sec. \ref{sec:real-catalog} , due to the
 magnitude limit of SDSS, it is difficult to find many clusters beyond
 $z\sim$0.4 . 
On the other hand, if we move the color cut bluer, we increase the contamination 
from $z\sim$0.3 galaxies (which are well within the magnitude limit of SDSS).
This is how the color cut box was optimized.

%
%
%


\subsection{Enhancement Method}\label{sec:enhance}

 After applying the color cuts, we use a special enhancement method
 to enhance the signal to noise ratio of clusters further.
First, we find all pairs of galaxies within five arcmin, this scale
 corresponds to the size of galaxy clusters at $z\sim$0.3 .
 Selecting larger separations blurs high $z$ clusters,
 while smaller separations  weaken the signal of low $z$ clusters.
 We empirically investigated several separations 
and found 5' to be a good mean value.
We then calculate the angular distance and color difference of each pair of galaxies.
We distribute a Gaussian cloud around the center position of each pair.
 The width of a Gaussian cloud is the angular separation of the pair and
 the volume of a Gaussian cloud is given by its weight ($W$), 
 which is calculated as:

%

\begin{equation}\label{equation}
 W =  \frac{1}{\Delta r + 1''} \times \frac{1}{\Delta (g^*-r^*)^2 +  2.5\times10^{-3}}
\end{equation}
,where $\Delta r$ is the angular separation between the two galaxies and 
$\Delta$($g^*-r^*$) is their color difference. Small softening parameters
(empirically determined)
are added in the denominator of each
term to avoid values becoming infinity.  
 This enhancement method provides stronger weights to pairs which
are closer both in angular space and in color space,
 thus are more likely to appear in galaxy clusters. 
Gaussian clouds are distributed in 30''$\times$30'' cells on the sky.
The 30'' cells are small compared to sizes of galaxy
clusters (several arcmins at $z\sim$0.5) .

\begin{figure}
\includegraphics[scale=0.4]{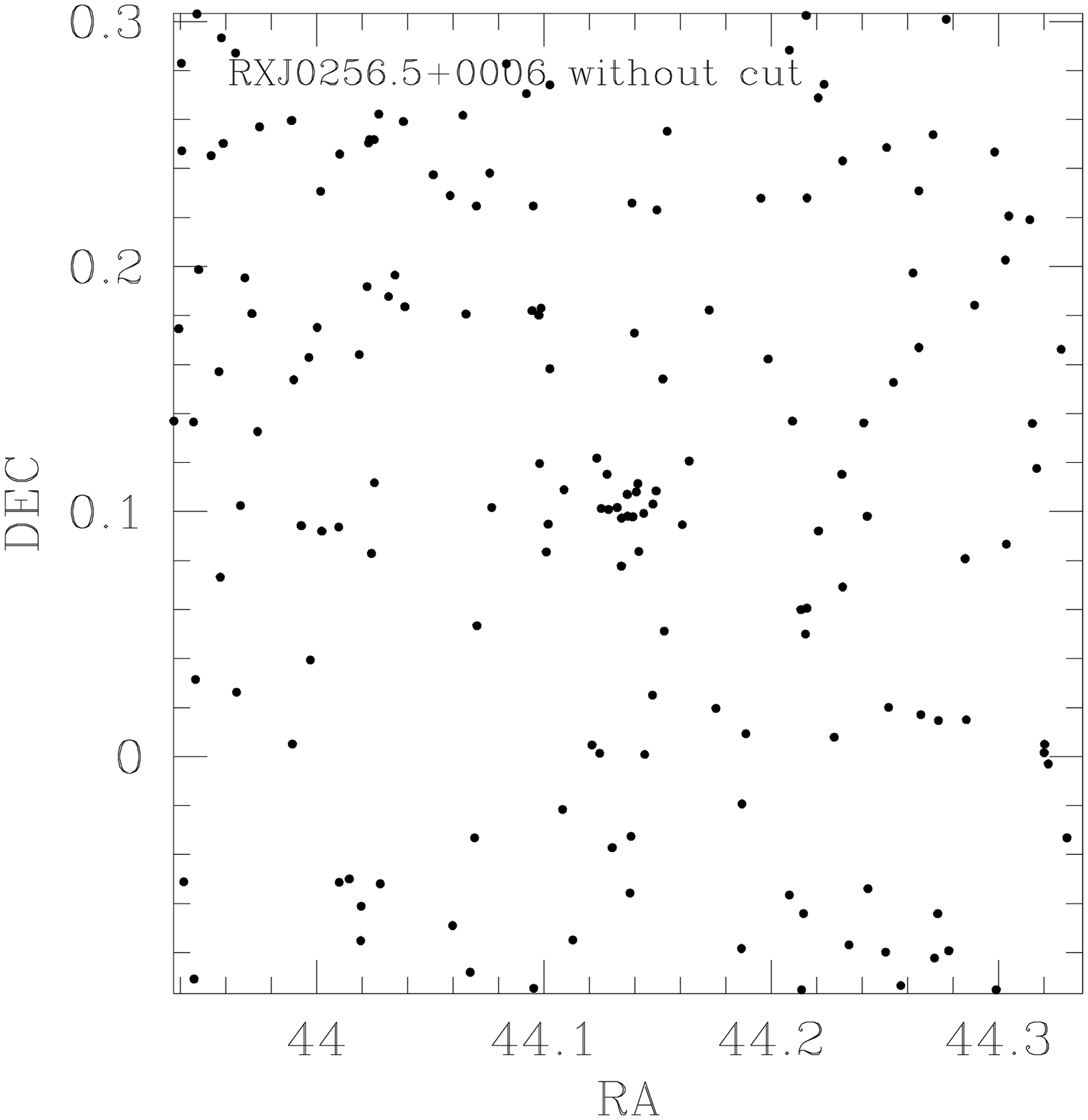}
\includegraphics[scale=0.4]{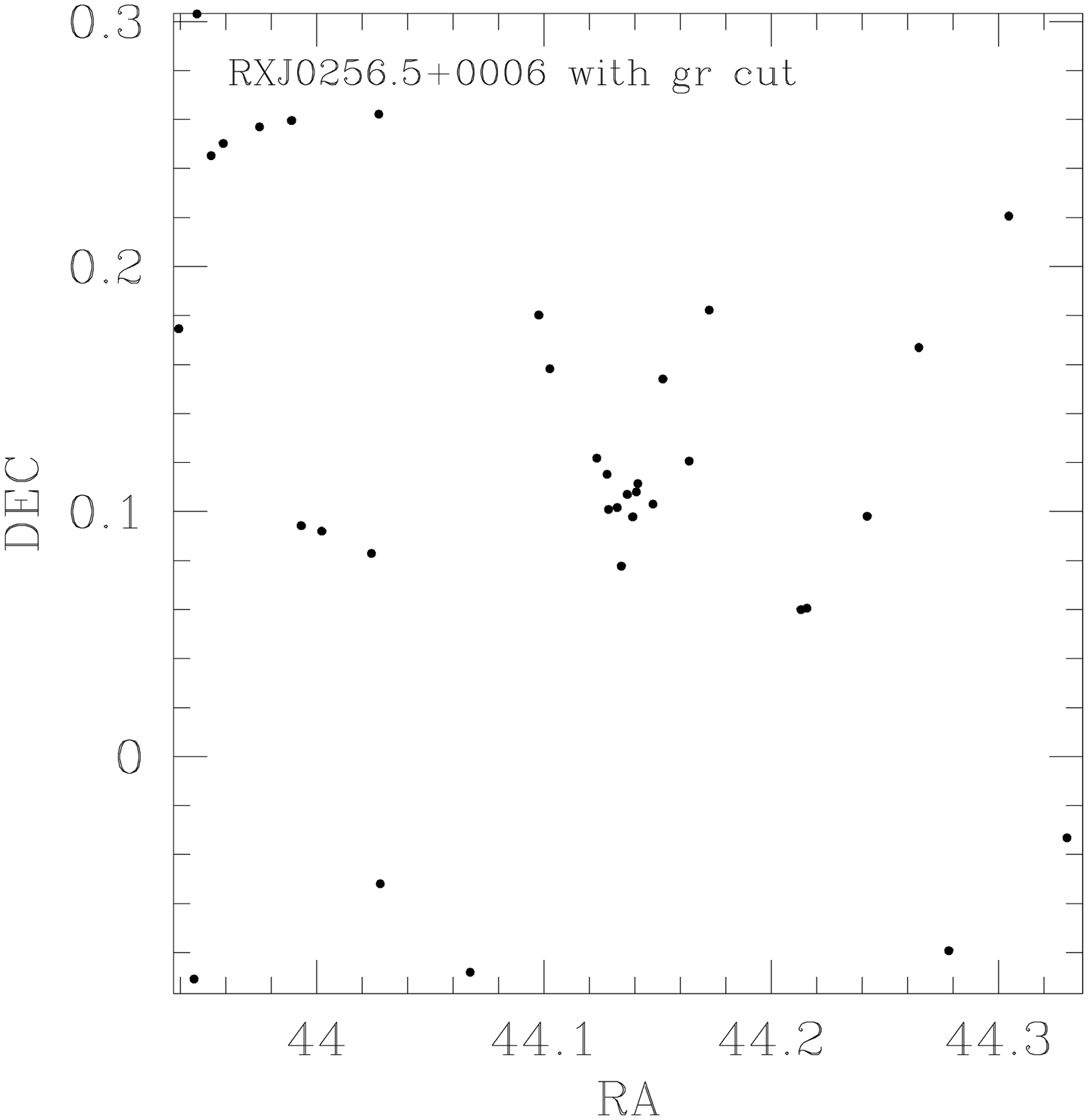}
\centering{\includegraphics[scale=0.5]{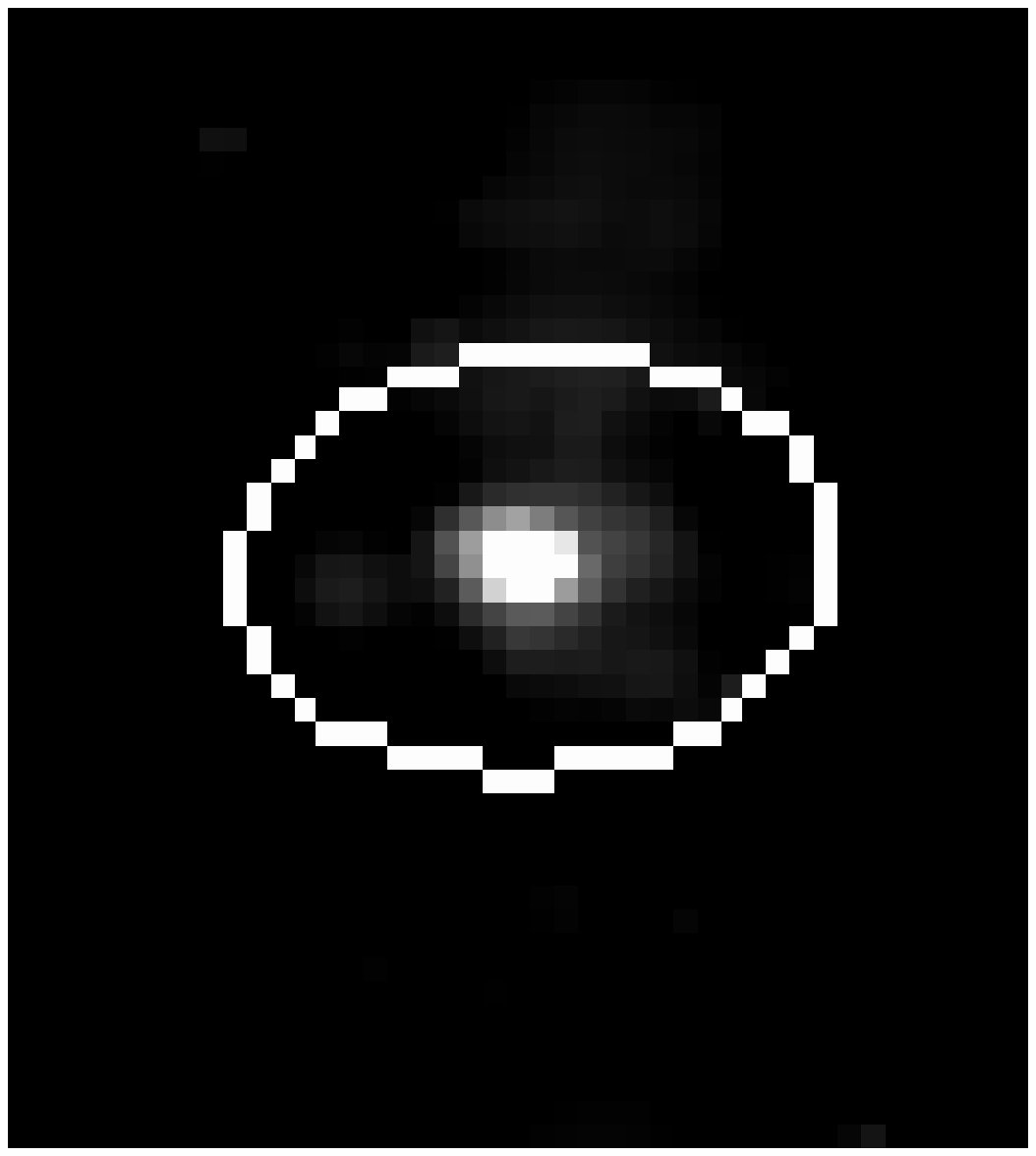}}
\caption{
\label{fig:beforecut_RXJ}
The distribution of galaxies brighter than $r^*$=20.0 around
 RXJ0256.5+0006. The upper left panel shows the distribution before
 applying any cut. The upper right panel shows the distribution after applying
 $g^*-r^*$ color cut. The lower panel shows the enhanced map.
 The color cut removes foreground and background galaxies as designed.
 RXJ0256.5+0006 is successfully detected as circled with a white line.
}
\end{figure}

\begin{figure}
\includegraphics[scale=0.4]{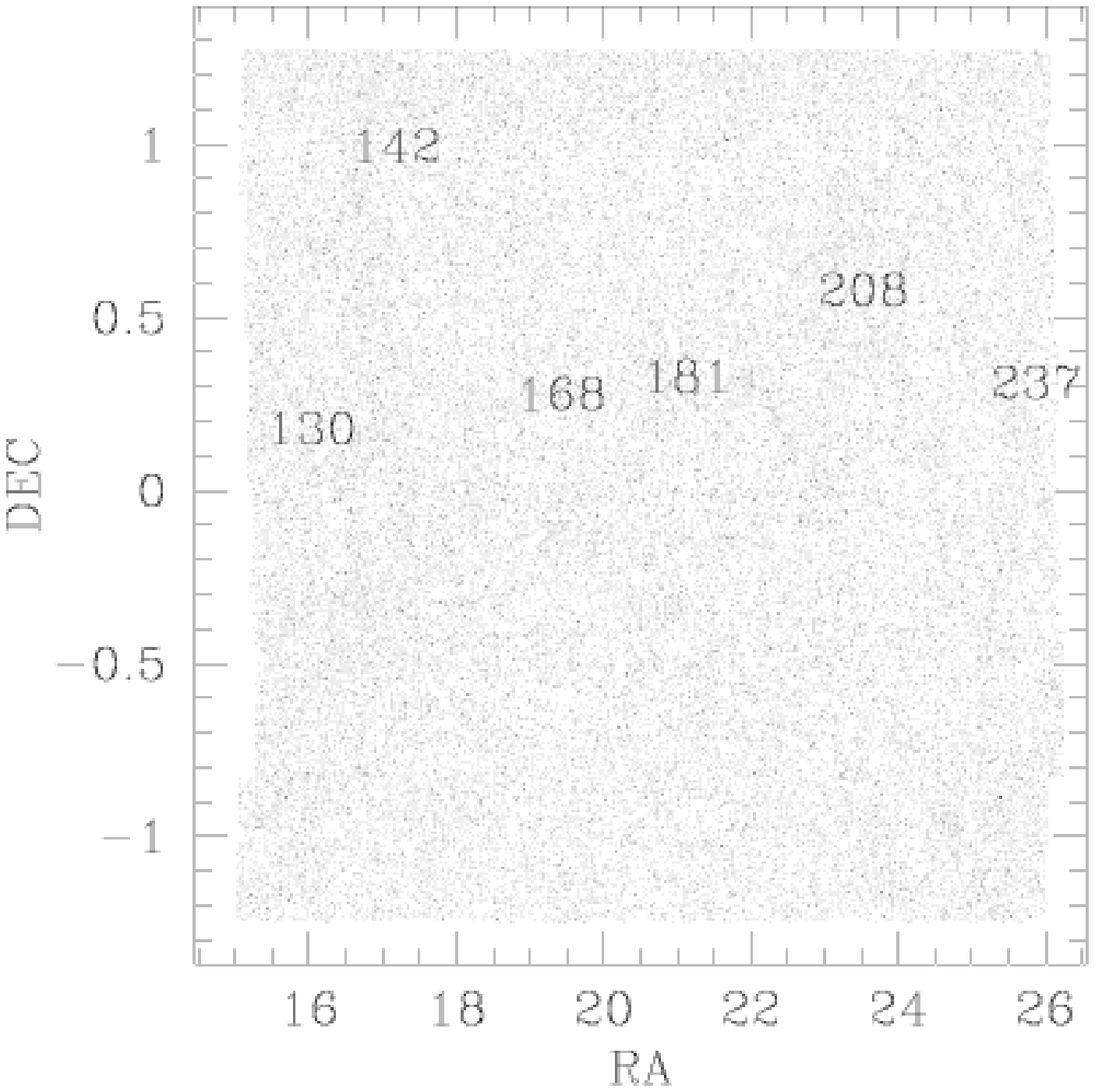}
\includegraphics[scale=0.4]{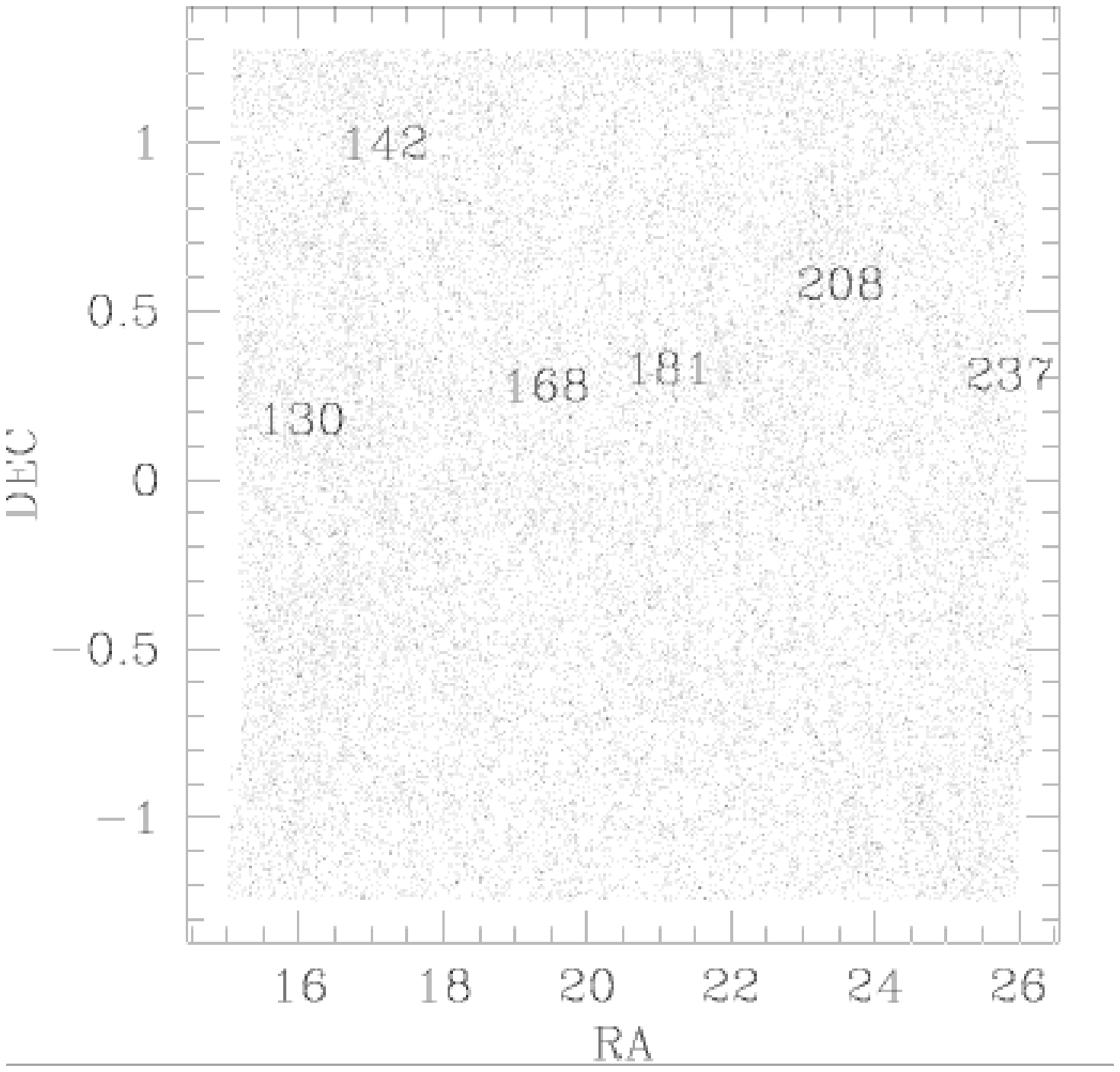}
\centering{\includegraphics[scale=0.5]{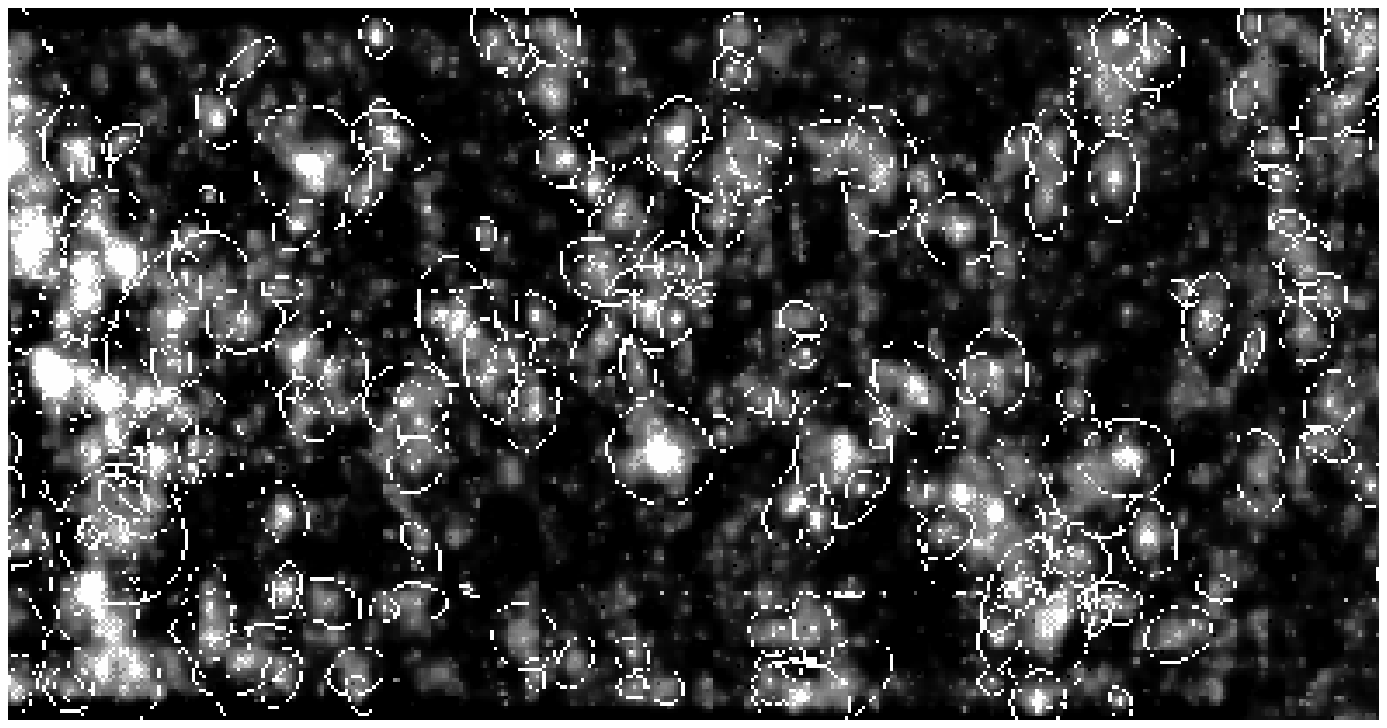}}
\caption{
\label{fig:beforecut_comparison}
The distribution of galaxies brighter than $r^*$=21.5.
The upper left panel shows the distribution before applying any cut.
The upper right panel shows the distribution after applying
 $g^*-r^*-i^*$ color-color cut. The numbers show the positions of Abell clusters.
 The lower panel shows the enhanced map in $g^*-r^*-i^*$ color-color
 cut. Detected clusters are circled with white lines.
}
\end{figure}

 An enhanced weighted map of high density regions is obtained by summing
 up the Gaussian clouds.
 The lower panels in Fig. \ref{fig:beforecut_RXJ} and
 \ref{fig:beforecut_comparison} present such enhanced maps of
 the region in their upper panels.
 RXJ0256.5+0006 is successfully enhanced in
 Fig. \ref{fig:beforecut_RXJ}.  Fig.  \ref{fig:beforecut_comparison}
 illustrates how the CE method finds galaxy clusters. 
 The advantage of this enhancement method in addition to the color
 cuts is that it makes full use of color concentration of
 cluster galaxies. 
 The color cuts are used to reduce foreground and background 
galaxies and to enhance the signal of clusters. 
Since the color-magnitude relation of cluster galaxies is frequently tighter than the width
of our color cuts, the use of the second term in equation (\ref{equation})  
- the inverse square of the color difference - further enhances the signal of cluster,
 in spite of the larger width of the color cuts.
 Another notable feature is that the enhancement method is adaptive. {\it
 i.e.} Larger separation pairs have large gaussian and small separation
 pairs have sharp, small gaussian. In this way, the enhancement method
 naturally fit to the any region with any number density of 
galaxies in the sky. It is also easy to apply it to data from another 
telescope with different depth and different galaxy density.
 Another benefit of the enhancement method is that it includes 
a smoothing scheme and thus 
conventional detection methods  commonly used in astronomical community  can be used to detect clusters in the
enhanced map.
The enhancement method uses the angular separation in the $W$. 
This might bias our catalog against nearby clusters ($z<$0.1),
 which have a large angular extent (and thus are given less $W$).
 However these nearby clusters
 already well documented in existing
catalogs;
 these nearby clusters will also be well sampled in the SDSS
spectroscopic survey with fiber redshifts, 
 and will thus be detected in the SDSS 3D cluster selection.
 (
Cut \& Enhance cluster detection method is
intended to detect clusters using imaging data only).
These nearby clusters do not have a significant effect on angular or
redshift-space correlations because the number of such clusters is a
small fraction of any large volume-limited sample.

\subsection{Detection}\label{sec:detection}


 We use SourceExtractor (Bertin et al.\ 1996) to detect clusters
 from the enhanced map discussed in Sect.\ref{sec:enhance}.  
 SourceExtractor identifies high density peaks  above a given threshold
 measuring the background and its fluctuation locally.
 The threshold selection determines the number of clusters obtained.
 A high threshold selects only the richer clusters.
 We tried several thresholds, examining the
 colored image, color-magnitude and color-color diagrams of  the
 resulting cluster catalog. The effect of changing threshold is 
 summarized in Table \ref{tab:sigma_test}. The numbers of clusters
 detected are not very sensitive to the threshold\footnote{The numbers
 of detection go up and down with increasing sigma because the following two effects cancel
 out each other. 1, Lower threshold detects faint sources and thus
 increases the number of detections. 2, Higher threshold deblends the peaks and
 increases the number of the detections.}.  
 Based on the above, we have selected the threshold to be 
 six times the background
 fluctuation, it is  the threshold which yields a large number of clusters
while the spurious detection rate is still low.  

 Monte Carlo simulations are sometimes used  to decide the optimal threshold,
 where most true clusters are recovered while
 the spurious detection rate is still low.
   However the simulations reflect an ideal situation, and they are 
inevitably different from true data; for example, a uniform background
 cannot represent the true galaxy distribution with its large scale
  structure. There are always clusters which do not match the
   radial profile or luminosity function assumed in Monte Carlo
   simulations 
 and this may affect the optimization of the threshold.
 The optimal threshold in Monte Carlo simulation differs from
 the optimal threshold in the real data.
 Therefore, we select the threshold empirically using the actual data and later
 derive the selection function using Monte Carlo simulation.

At high redshifts ($z>$0.4), the number of galaxies within the color
cuts is small; therefore the 
 rms of the enhanced map is generally too low and
the clusters detected at high redshift have
 unusually high signal. 
 To avoid such spurious detections,
we applied another threshold at maximum absolute 
flux=1000\footnote{FLUX\_MAX+BACKGROUND=1000, where FLUX\_MAX and
 BACKGROUND are the parameters of Source Extractor.
  FLUX\_MAX+BACKGROUND is the highest
 value in the pixels within the cluster.  It is an absolute value, and 
not affected by rms value.} in the enhanced map.
Spurious detections with high signal would generally have low values
because they are not true density peaks.
 The maximum absolute flux=1000 threshold can thus reject spurious detections.
The value is determined by investigating the image, color-magnitude and
 color-color diagrams 
 of the detected clusters and iterating the detection with different values of
 the maximum absolute flux threshold. The effect of changing the
 absolute flux threshold is summarized in Table  \ref{tab:fluxmax}

 To secure the detection further, at all redshifts, we demand at least
 two detections in the 34 cuts. 
 This is demanded because the cluster galaxies have color concentrations 
 in all $g^*-r^*$, $r^*-i^*$ and $i^*-z^*$ colors; 
 real clusters should thus be
 detected in at least two color cuts.


\subsection{Merging}
 We apply the procedure of cut, enhance and detect to all of the 34 color
cuts (30 color cuts + four color-color cuts) independently. 
After creating the 34 cluster lists, we 
merge them into one cluster catalog. 
We regard the detections within 1.2 arcmins as one cluster.
 To avoid two clusters with different redshifts being merged into one cluster
due to the chance alignment,
we do not merge clusters that are detected in two color cuts of the same bands 
 unless the successive color cuts both detect it.

 An alternative way to merge clusters would be 
 to merge only those clusters which are detected 
in the consistent color cut in all $g^*-r^*$, $r^*-i^*$ and $i^*-z^*$ colors,
 using 
 the model of the elliptical galaxy colors.
 However, the catalog will be biased against clusters which have
 different colors than the model ellipticals.
 In order to minimize the assumptions on cluster properties 
 we treat the three color space, $g^*-r^*$, $r^*-i^*$ and $i^*-z^*$ , independently.

\subsection{Redshift and Richness Estimation}\label{sec:real-catalog}
 We estimate the redshift and richness of each cluster as
 follows.
 In stead of the same richness estimator as Abell's,
 we count the number  of galaxies inside the detected cluster radius 
which lie in the two magnitude range ($r^*$ band) from $m_3$ (the third
 brightest galaxy) to $m_3$+2 (CE richness).
 The difference from Abell's estimation is that he used a fixed 1.5 $h^{-1}$Mpc as
a radius. 
 Here we use the detection radius of the cluster detection algorithm
 which can be larger or smaller than
Abell radius, typically slightly smaller than  1.5 $h^{-1}$Mpc.
 The background galaxy count is subtracted using the average galaxy counts
in the SDSS commissioning data.

 For the redshift estimates, we use 
the strategy of the redshift estimation of the maxBCG technique
(Annis et al.\ 2001). 
 We count the number of galaxies within the detected radius 
 that are brighter than $M^*_{r^*}$=-20.25 for a given redshift assumed
 and are within a color range of $\pm$1 mag in $g^*-r^*$ around the color
 prediction for elliptical galaxies  (Fukugita et al.\ 1995).
 This is determined in estimated redshift step of $\delta z$= 0.01.
 After subtracting average background number counts from each bin,
the redshift of the bin that has the largest number of galaxies is
taken as the  estimated cluster redshift. 
The estimated redshifts are calibrated using the spectroscopic redshifts
from the SDSS spectroscopic survey. Our redshift estimation depends on
the model of Fukugita et al. (1995), but the difference from other
models are not so significant, as  seen in 
the difference between open triangles (PEGASE model) and
filled triangles (Fukugita et al. 1995) of Fig. \ref{fig:color-cut-test}.
If a cluster has enough elliptical galaxies, the redshift of the cluster
is expected to be well measured.
If a cluster is , however, dominated only by spiral galaxies, as seen in
the difference between triangles and squares, the redshift of the
cluster will be underestimated.

\begin{figure}
\begin{center}
\plotone{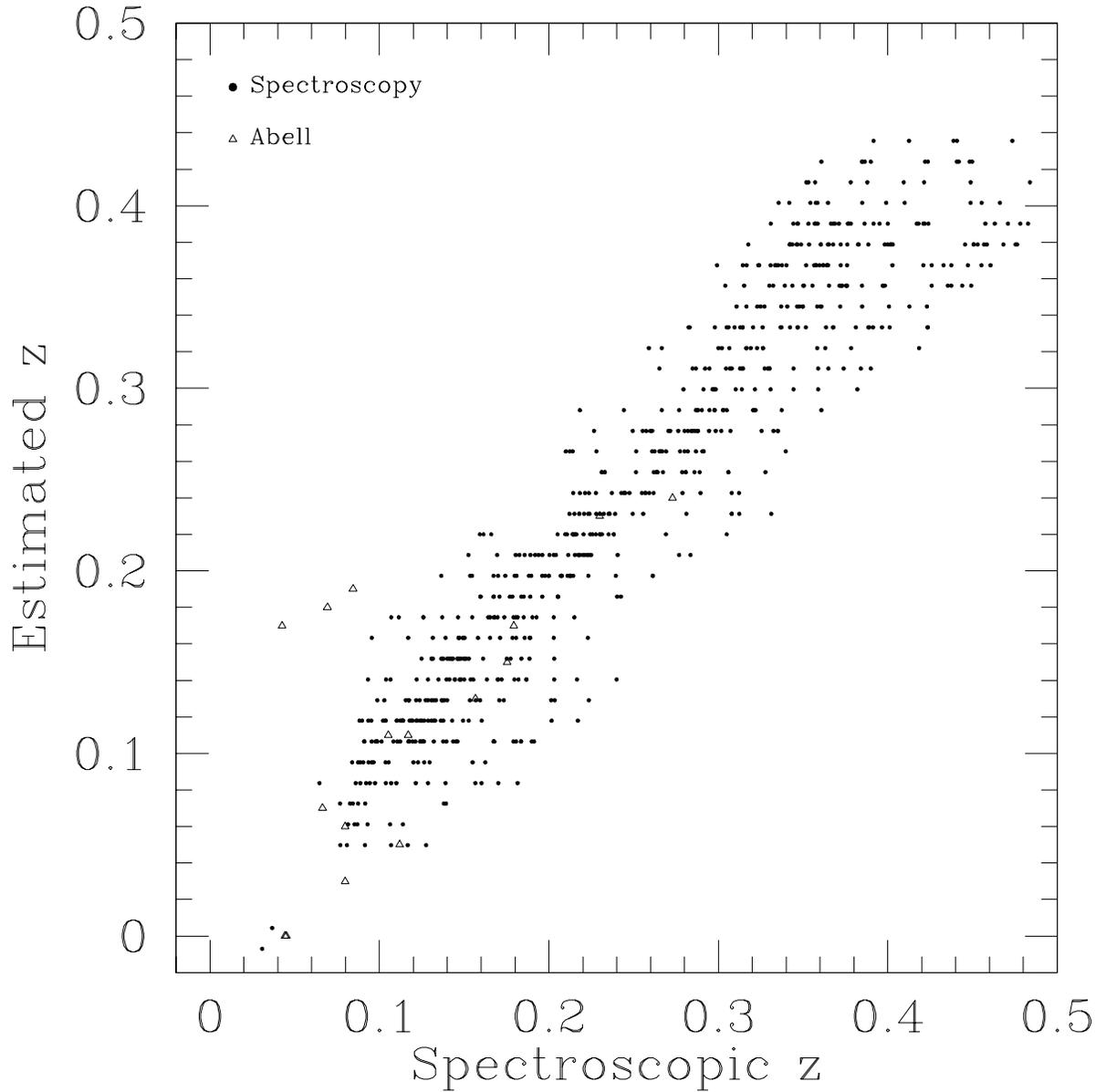}
 \caption{
 \label{fig:zaccuracy.eps}
 The redshift estimation accuracy. 
 The estimated redshifts are plotted against spectroscopic redshifts.
 Abell clusters are plotted with triangles.
 Dots are the redshifts from SDSS spectroscopic galaxies.
 Extensive outliers $\delta z>$0.1 are removed.
 The dispersion is 0.0147 for $z<$0.3 and 0.0209 for $z>$0.3 .
}
\end{center}
\end{figure}

 Fig.\ \ref{fig:zaccuracy.eps} shows the redshift accuracy of the method.
 The estimated redshifts are plotted against observed redshifts from the
 spectroscopic observation.
 The redshift of the SDSS spectroscopic galaxy within the detected
radius and with nearest spectroscopic redshift to the estimation is
adopted as the real redshift. In the fall equatorial region,
 699 clusters have spectroscopic redshifts.
 The correlation between true and estimated redshifts is very good:
 the rms scatter is $\delta z$=$\pm$0.0147 for $z<$0.3
clusters, and $\delta z$=$\pm$0.0209 for $z>$0.3 clusters.
 Triangles show 15 Abell clusters measured with available
 spectroscopic redshifts,
 there are three outliers at low
 spectroscopic redshifts. CE counterparts for these three  clusters all have
 very small radii of several arcmin. Since these Abell
 clusters are at $z<$0.1, 
 the discrepancy is probably not in the redshift estimation 
 but rather in the detection radius.
 We construct the SDSS Cut \& Enhance galaxy cluster catalog containing
 4638 galaxy clusters.
 The catalog is available at the
 following website. http://astrophysics.phys.cmu.edu/$\sim$tomo


\section{Monte Carlo Simulation}\label{sec:monte}
 In this section, we examine the performance of the Cut \& Enhance method and determine the selection function using Monte Carlo simulations. We also perform false positive tests.

\subsection{Method}

 We perform Monte Carlo simulations both with a real background using the SDSS
 commissioning data and with the shuffled background. (We explain these below.)
 For the real background, we randomly choose a 1 deg$^2$ region of the SDSS
 data with seeing better than 1.7''.\footnote{
 Though the SDSS survey criteria for seeing
is better than 1.5'', some parts of the SDSS commissioning data have seeing worse than 2.0''.
It is expected that the seeing is better than
1.5'' for all the data after the survey begins.}
 For the shuffled background,
 we re-distribute all the
 galaxies in the above 1 deg$^2$ of SDSS data randomly in position , but keep
their colors and magnitudes unchanged.

 Then, we place artificial galaxy clusters on these backgrounds.
 We distribute cluster galaxies randomly using a King profile (King
 1966; Ichikawa
 1986) for the  radial density, 
 with concentration index of 1.5 and cut
 off radius of $2.1h^{-1}$Mpc, which is the size of Abell 1577 (Struble \& Rood  1987). 
 For colors of the artificial cluster galaxies, we use the color and magnitude
distribution of Abell 1577 (at $z\sim0.14$, Richness$\sim1$) as a model.
 We choose the SDSS fields
which cover the entire Abell 1577 area and count the number of galaxies in
each color bin. The size of the bins is 0.2 magnitude in both colors and
magnitude. 
The color and magnitude distribution spans in four dimension space, $r^*$,
$g^*-r^*$, $r^*-i^*$ and $i^*-z^*$.
 We count the number of field galaxies using the same size
 fields near Abell 1577 and subtracted the distribution of field
galaxies from the distribution of galaxies in the Abell 1577 fields. 
 The resulting 
color distribution is used 
as a model for the artificial galaxy clusters.
 Galaxy colors are selected randomly so that they reproduce the overall color
distribution of Abell 1577. 
 The distribution is linearly interpolated when allocating colors and
magnitudes to the galaxies.

%
%


 For the high redshift artificial clusters, we apply k-correction and
 the color prediction of elliptical galaxies from Fukugita et al.\
 (1995).
  For the color prediction, only  the color difference, not the absolute
 value, is used.
Galaxies which become fainter than $r^*$=21.5 are
 not used in the  Cut \& Enhance method.

%
%
%
%

\subsection{Monte Carlo Results}

 First, we run a Monte Carlo simulation with only the background,
  without any
 artificial clusters, in order to measure the detection rate of the
simulation itself.
The bias detection rate is defined as the percentile 
in which any detection is found within 1.2 arcmins from the
position where we later place an artificial galaxy cluster.
The main reason for the false detection is 
that a real cluster sometimes comes into the detection 
position, where an artificial cluster is later placed.
 This is not the false detection of the Cut \& Enhance method but rather
  the noise
 in the simulation itself.
The bias detection rate with the real SDSS background is
4.3\%.
 This is small relative to the actual cluster detection rate discussed below.
 The bias detection rate using the shuffled background is lower, as expected.
 It is 2.4\%. 
 We run Monte Carlo
 simulations with a set of artificial clusters with redshifts ranging 
 from $z$=0.2 to $z$=0.6, and with richnesses of $Ngal$= 40, 60, 80
 and 100, at each redshift. ($Ngal$ is the number of galaxies
 within $2.1h^{-1}$Mpc inputted
 into a cluster, whose magnitudes are $r^*<$21.5 at the redshift of A1577.)
 If a galaxy becomes fainter than $r^*=21.5$, it is
not counted in the Cut \& Enhance detection method
 even if it is included in $Ngal$).
For each set of parameters, the simulation is iterated
1000 times. 

\begin{figure}
\begin{center}
\plotone{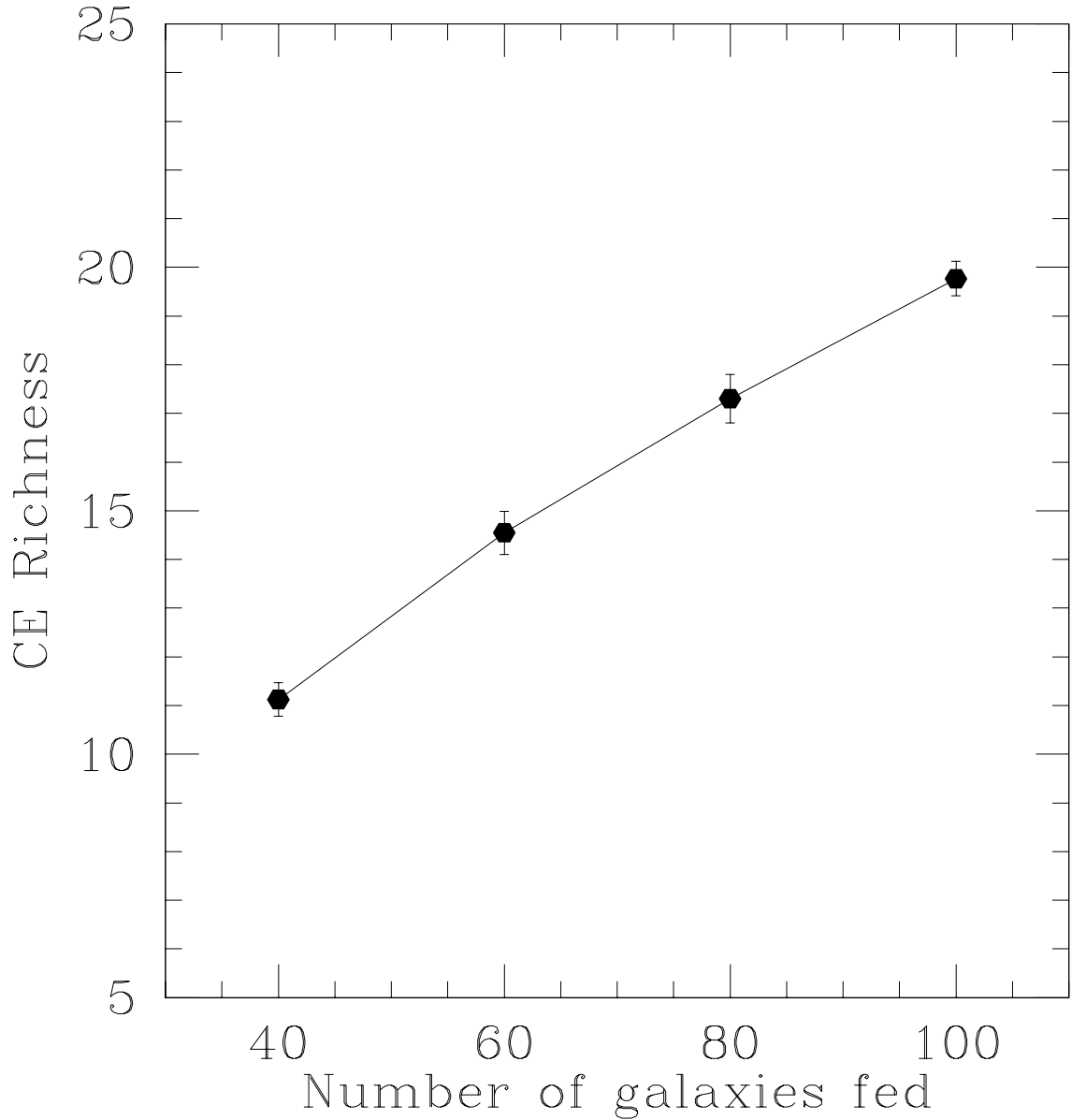}
\caption{
\label{fig:ngal-ab1ellrich}The number of galaxies fed ($Ngal$) v.s. richness. The number of galaxies put in
 the artificial cluster is compared with richness (the number
 of galaxies within the detected radius whose magnitude is between the magnitude of the third
 brightest galaxy and 
the magnitude fainter by two). The error bars show 1$\sigma$ standard error.}
\end{center}
\end{figure}

 In Fig.\ \ref{fig:ngal-ab1ellrich}, we compare $Ngal$ 
 with cluster richness
 where richness is defined (Sect.\ref{sec:method}) as the number of galaxies 
 within the two magnitude range below the third brightest galaxy, located within the
 cluster radius that the Cut \& Enhance method returns. 
 The error bars are 
 $1\sigma$ standard  error. $Ngal$=50 corresponds to  Abell richness
 class $\sim$1.

\begin{figure}
\begin{center}
\plotone{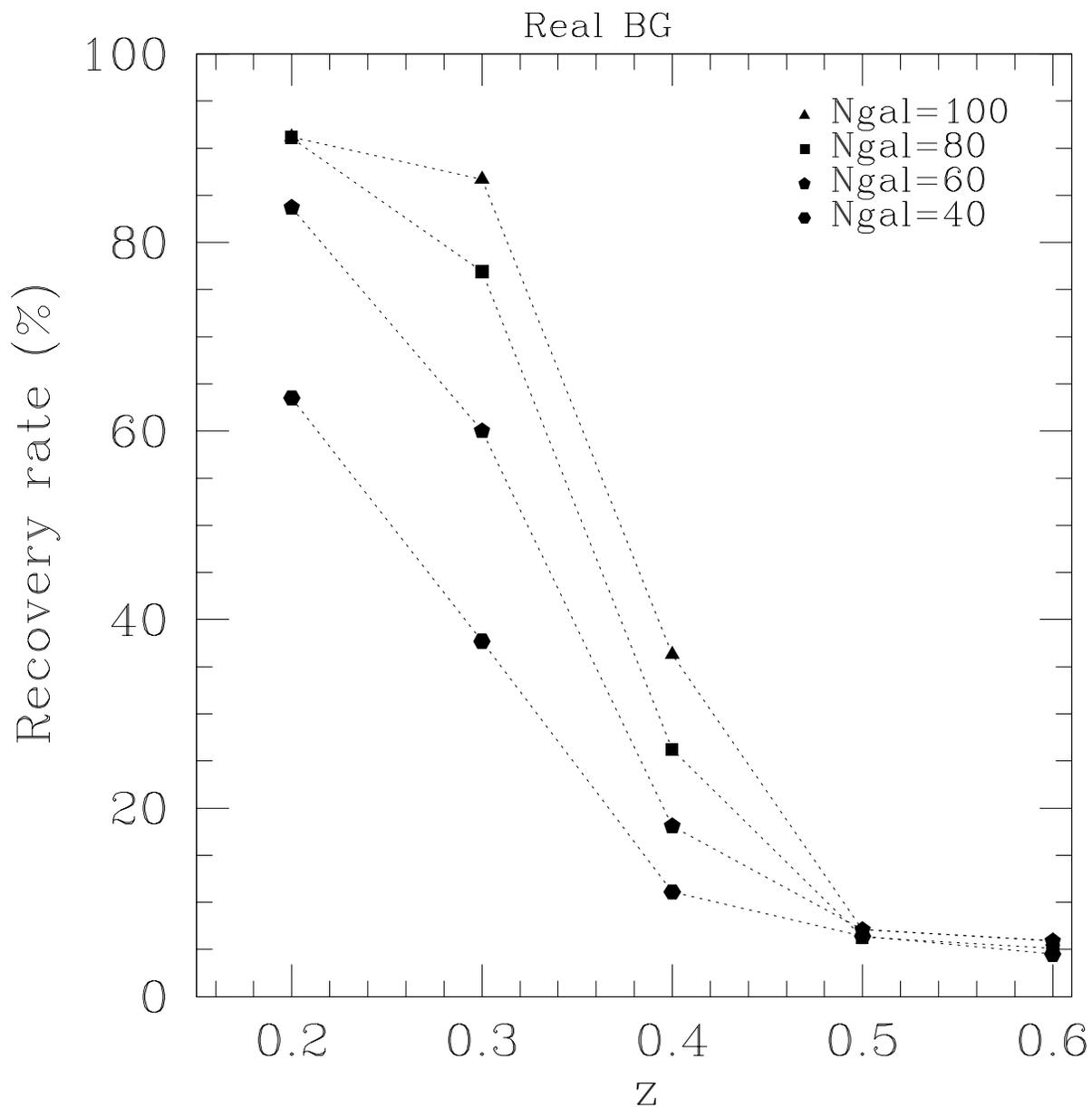}
\caption{
\label{fig:monte-recovery-simu}
Recovery rate in Monte Carlo simulation with the real SDSS background.
Recovery rate is plotted against redshift.
The artificial clusters are added on the real SDSS background randomly
 chosen from the SDSS commissioning data. The detection is iterated 1000
 times for each data point. Even at $z$=0.5, $Ngal$=50 cluster is detected with more
 than 82.5\% 
probability.
}
\end{center}
\end{figure}

Fig.\ \ref{fig:monte-recovery-simu} shows the recovery rate in the Monte
Carlo simulations 
on the real background.
 The percentage recovery rate is shown as a function of redshift.
Each line represents
 different richness input clusters, $Ngal$=100, 80, 60 and 40, from top to
bottom.
Because the false detection rate in the simulation with real
background is 4.3\%, all the lines converges to 4.3\% at high redshift.
 The detection rate drops suddenly at $z$=0.4 because  At this point, a
 large fraction of the cluster member galaxies are lost  
 due to the magnitude limit of $r^*=21.5$. Roughly speaking, it
 determines the depth of a SDSS cluster catalog.
 The $Ngal$=80  clusters are recovered 
$\sim$80\% of the time to $z<$0.3 dropping to $\sim$40\%
 beyond  $z\sim$0.4.
 Clusters of the lowest richnesses, 
 $Ngal$=40 clusters are more difficult to detect, as expected.
The detection rates of $Ngal$=40 clusters are  
less than 40\% even at $z$=0.3.
The recovery rate for $Ngal$=100 at $z=0.2$ is not 100\%.
 If we widen the detection radius from 1'.2 to 5'.4, the recovery rate
 increases to 100\%.  
 Note that the radius of 5'.4 is still small in comparison with
the size of A1577: 11' at $z$=0.2 (Struble \& Rood 1987).
 The reason may be that a real cluster (in the real background) 
 is located close to the artificial cluster, and the detected
 position may then be shifted by more than 
 the detection radius (1'.2) away from the cluster.

\begin{figure}
\begin{center}
\plotone{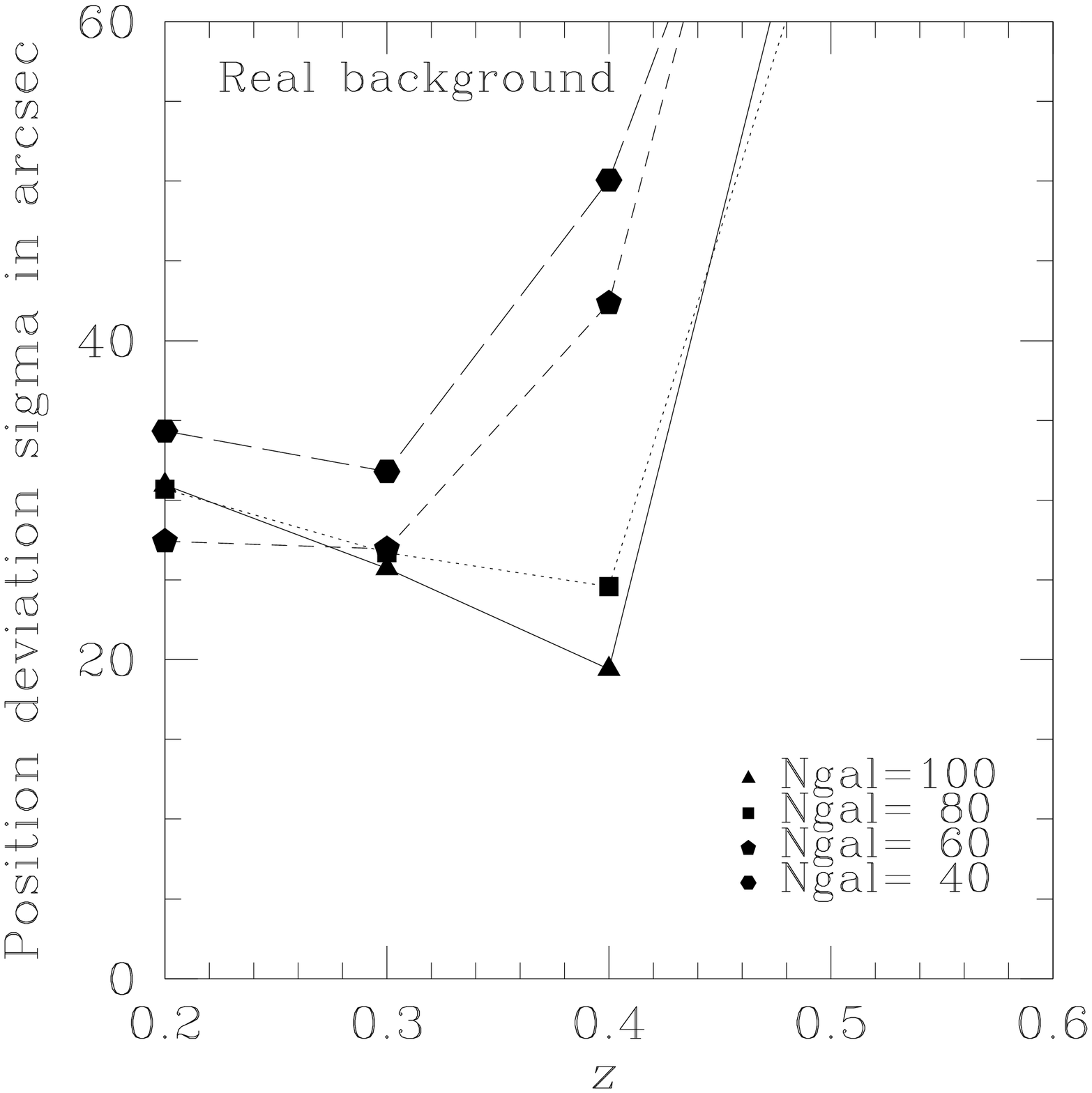}
\caption{
\label{fig:z-posi-simu}
Positional accuracy with the real SDSS background. The positional accuracy is
 almost constant because the more distant cluster is more compact in
 angular space. Positional accuracy of $\sim$0.01 deg is reasonable considering
 that the mesh size of the enhancement method is 30''(=0.0083deg).
The lack of some points at low richness and high redshift is due to the
 failure to fit using poor statistics.
}
\end{center}
\end{figure}

Fig.\ \ref{fig:z-posi-simu} shows the positional accuracy of the detected
 clusters in the Monte Carlo simulation with the real SDSS
background, as a function redshift and richness.
 The 1$\sigma$ positional errors of the detected clusters is shown.
 Note that since CE does not detect much fraction of clusters beyond
 $z$=0.4, there is not much meaning in discussing the position accuracy
 of beyond  $z$=0.4.
 The positional accuracy is
 better than 1' until $z$=0.4 in all richness ranges used.
 The deviation is nearly independent of the redshift because the high redshift
 clusters are more compact than the low redshift ones. 
 This partially cancels the effect of losing  more galaxies at high redshift
 due to the flux limit of the sample.
 The positional accuracy roughly corresponds to the mesh size of the enhancement
 method, 30''.
 As expected, the positional accuracy is worse for high redshift poor
 clusters ($z$=0.4 and $Ngal\leq$60).
The statistics for these objects are also less good; 
 the detection rate of $Ngal$=60 and 40 clusters are less than 20\% at $z$=0.4.
%

\begin{figure}
\begin{center}
\plotone{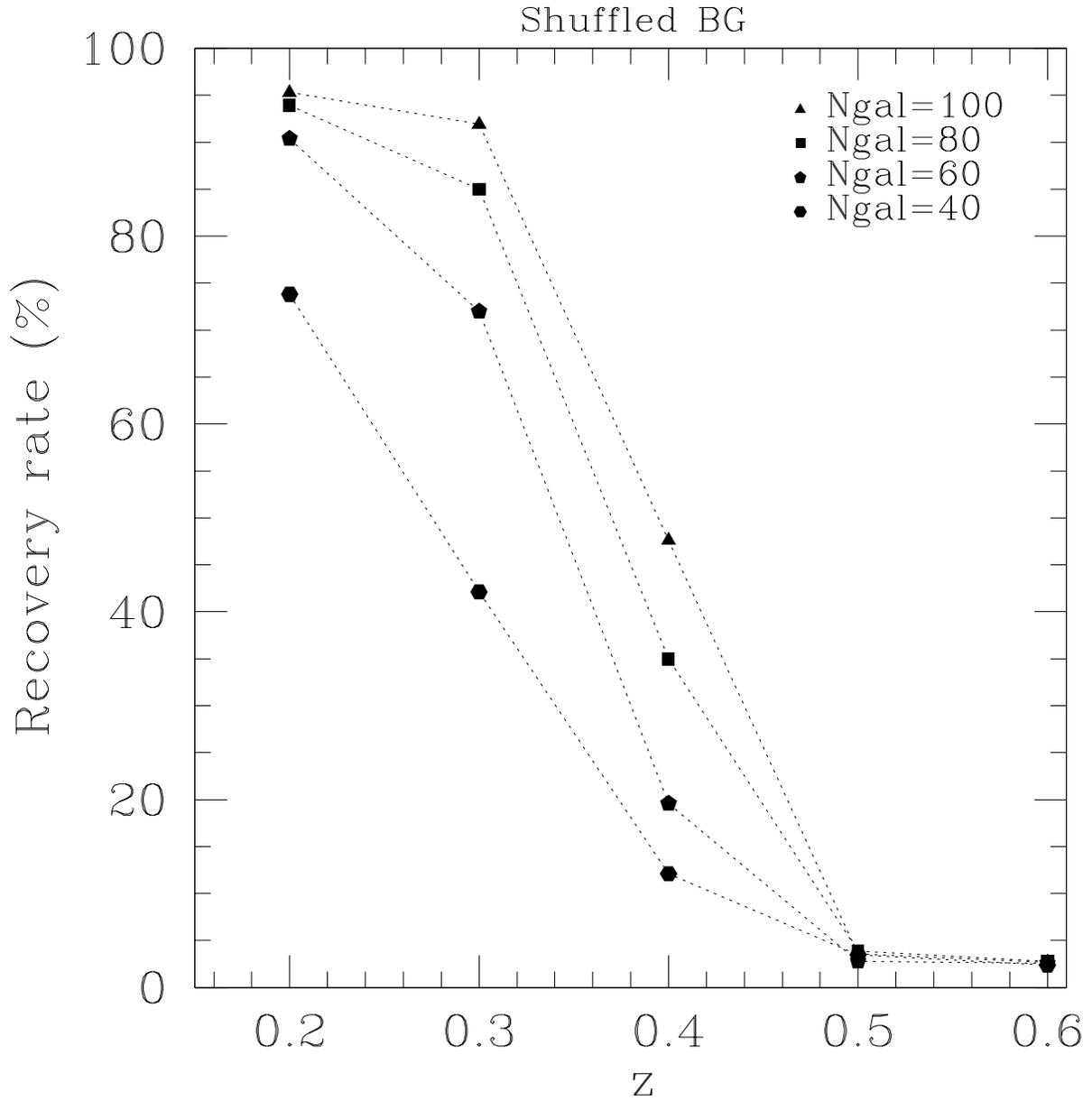}
\caption{
\label{fig:monte-recovery-uni}
Recovery rate in Monte Carlo simulation with the shuffled background.
The artificial clusters are added on the shuffled background randomly
 chosen from the SDSS commissioning data. The detections are iterated 1000
 times. Even at $z$=0.5, $Ngal$=50 cluster is detected with more than 80\% percentile.
}
\end{center}
\end{figure}

\begin{figure}
\begin{center}
\plotone{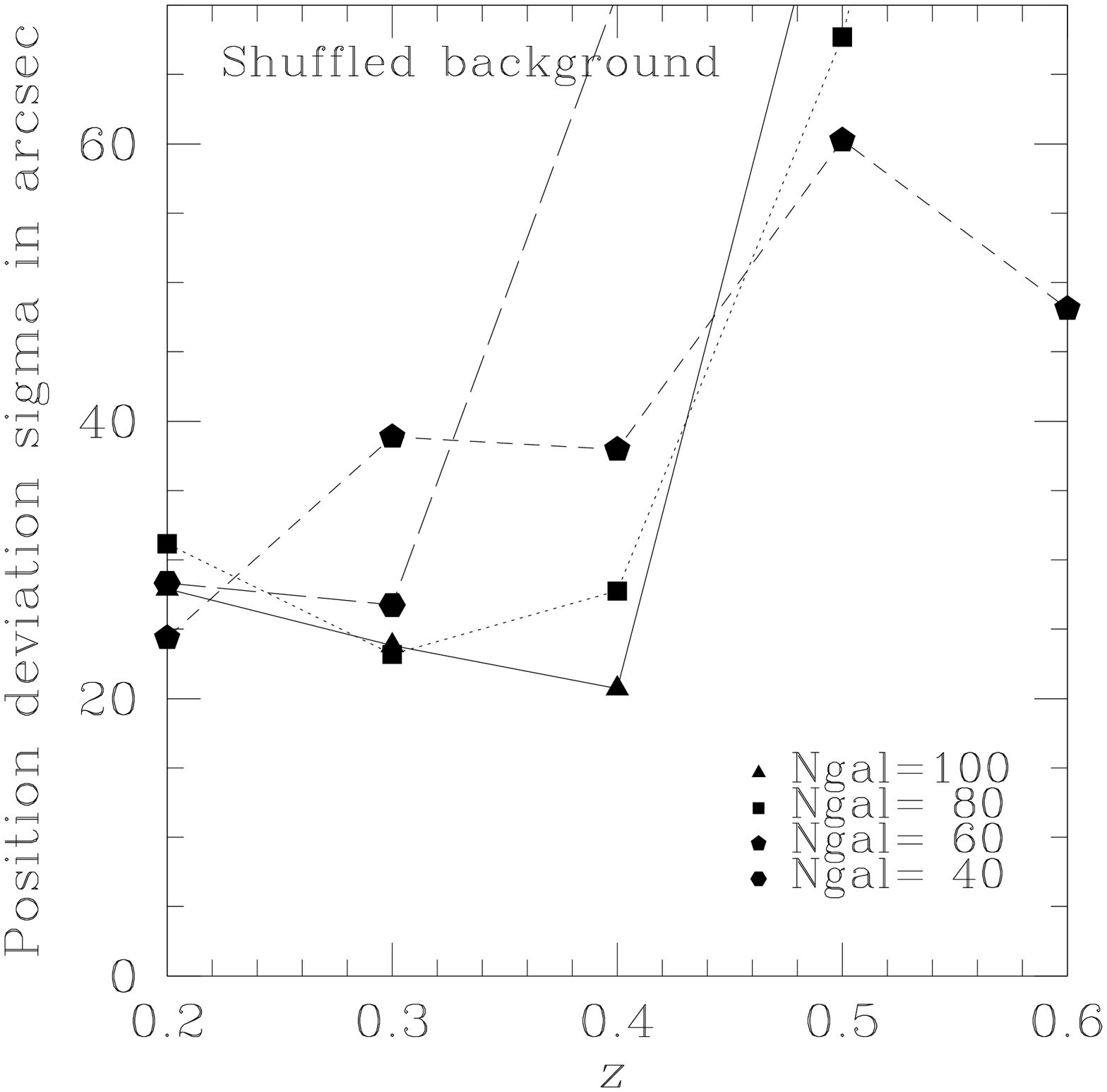}
\caption{
\label{fig:z-posi-uni}
Positional accuracy with the shuffled background. The positional accuracy is
 almost constant because the more distant cluster is more compact in
 angular space. Positional accuracy of $\sim$ 0.01 deg is good considering
 that the mesh size of the enhancement method is 30''(=0.0083deg).
The lack of some points at low richness and high redshift is due to the
 failure to fit using poor statistics.
}
\end{center}
\end{figure}

Fig.\ \ref{fig:monte-recovery-uni} presents the recovery rate of artificial
clusters in Monte Carlo simulations with the shuffled background. The
recovery rates are slightly better than with the real background. Again, the recovery rates drop sharply at $z$=0.4.
 The $Ngal$=100  clusters are recovered with $\sim$90\% probability to
 $z\sim$0.3 and $\sim$40\%  at $z\sim$0.4.
The detection rate is slightly higher than with the real background.
 At $z\leq$0.3,  $Ngal>$40 clusters are recovered at $>$40\%.
 Fig.\ \ref{fig:z-posi-uni} shows the positional accuracy of the detected
 clusters in the simulations (with shuffled background).
 The results are similar to these with the real data background.
 The positional accuracy is
 better than 40'' until $z$=0.3 for all richnesses.

\subsection{False Positive test}\label{sec:false-positive}

\begin{figure}
\begin{center}
\plotone{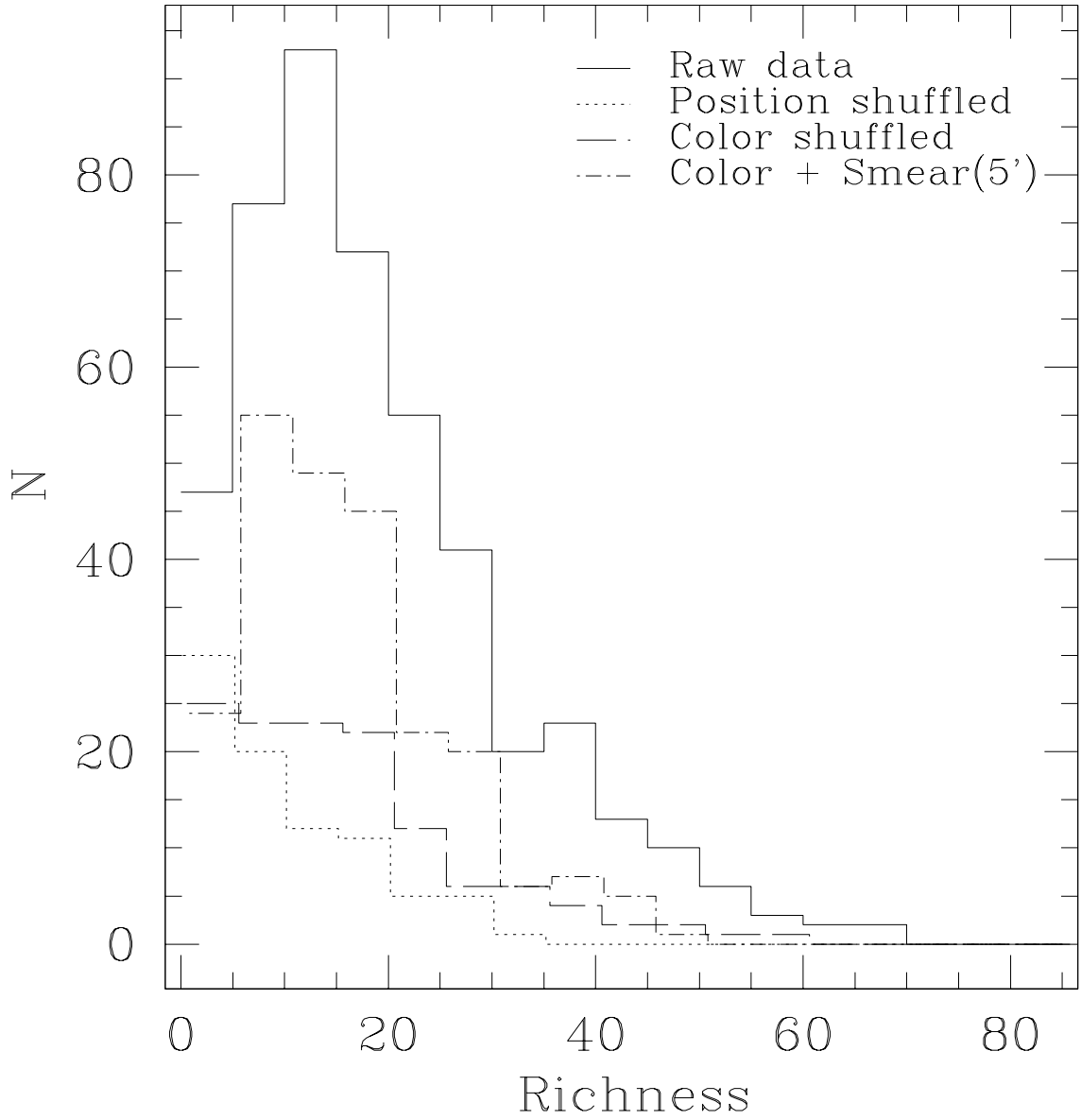}
\caption{
\label{fig:tim_test}
False positive test. Detection test is performed using 25 deg$^2$ of
 SDSS commissioning data. Solid line represents the results with real
 data. Dotted line represents the results with position shuffled
 data. Long dashed line is for color shuffled data subtracting the
 detection from the real data. Short dashed line is for color shuffled
 smearing data.}
\end{center}
\end{figure}

\begin{figure}
\begin{center}
\plotone{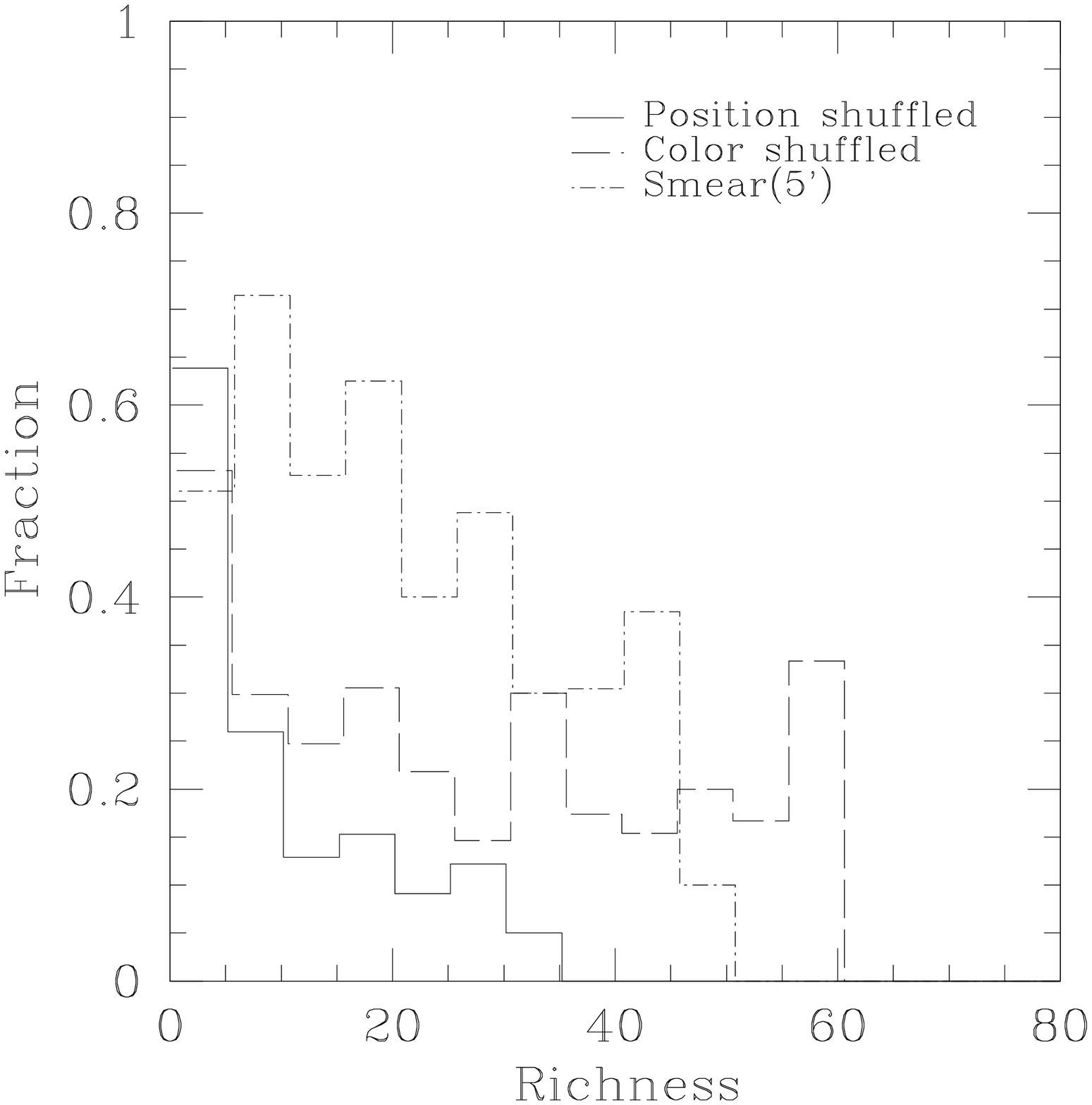}
\caption{
\label{fig:tim_test_ratio}
False positive test. Detection test is performed using 25 deg$^2$ of
 SDSS commissioning data. Each line represents the ration to the real
 data at the richness bin.
 Dotted line represents the results with position shuffled
 data. Long dashed line is for color shuffled data subtracting the
 detection from the real data. Short dashed line is for color shuffled
 smearing data.}
\end{center}
\end{figure}

 In order to test false positive rate, we prepared 
 four sets of the data: 
 1) Real SDSS data of 25 deg$^2$.
 2) Position of galaxies in the same 25 deg$^2$ are 
 randomized (galaxy colors untouched.) 
 3) Colors of galaxies are shuffled. (galaxy position untouched.) 
 4) Color is shuffled and position is smeared (5'). Galaxy colors are randomized 
 and positions are randomly distributed in the way that galaxies still lie within 5'
 from its
 original position. This is intended to include large scale
 structure without galaxy clusters.
 The results are shown in Fig. \ref{fig:tim_test}. 
  Solid line represents the results with real
 data. Dotted line represents the results with position shuffled
 data. Long dashed line is for color shuffled data. For color shuffled 
 data, we subtracted the detections in real data, because it still
 contains real clusters there. The fact that color shuffled data 
 still detects many clusters are consistent with the generous color cuts
 of CE.
 Short dashed line is for color shuffled
 smearing data.
 In Fig. \ref{fig:tim_test_ratio}, the ratio to the real data is plotted 
 against CE richness. 
 The promising fact is that not so many sources are detected from 
 position shuffled data. The rate to the real data is below 20\% at
 CE richness $>$20. 
More points are detected from color shuffled data 
 and smearing data but this does not mean the false positive rate of CE
 is as high as those values. 
Smearing data still has a
 structure bigger than 5', and they can be real clusters. 
 Overall, our simulations show that for richness $>$10, over 70\% of 
 CE clusters are likely to be real systems (as shown by the color \& position
 shuffled simulations.)

%

%
%
%

%
%



\section{Visual inspection}\label{sec:visual}

 To investigate  whether the detected clusters are true clusters or spurious
detection, spectroscopic observations are necessary.
 Although  large spectrometers which can observe the spectra of many
galaxies at one time are becoming available ($e.g.$ SDSS, 2dF),
it is still time consuming.
Since the SDSS Cut \& Enhance cluster  catalog 
will have more than 100,000 galaxy
clusters when the survey is complete, it is in fact impossible to 
spectroscopically confirm all the clusters in 
the SDSS Cut \& Enhance cluster catalog.
As a preliminary check of our method, 
we visually inspect all the Cut \& Enhance clusters within a given area
(right ascension between 16 deg and 25.5 deg and declination between
-1.25 deg and +1.25 deg, totaling  23.75 deg$^2$. The region in Fig.  \ref{fig:beforecut_comparison}).
 A total of 278 CE
galaxy clusters are located within this area (
after removing clusters touching the region's borders).
 Out of the 278 CE 
galaxy clusters, 
 we estimate that 
10 
are false detections.
 Since the strategy of the Cut \& Enhance method is to detect 
 every clustering of galaxies, 
 we call every angular clustering of galaxies with the same color  
a successful detection here. (As we show in Sect. \ref{sec:false-positive},
 30\% of clusters could be false detections, such as chance projections.)

 Among the 10 false detections, three are  bright big galaxies deblended
into several pieces. 
In the other cases,
a few galaxies are seen but not an apparent cluster or group.
 (In one case a rich cluster exists
about six arcmin from the false detection).
The 10 false detections are summarized in Table \ref{tab:ce-miss-new}.
 $\sigma$ (column[1]) is the significance of the detection; 
CE richness (column[2])  is its richness;
$z$ (column[3]) is the color estimated redshift;
 and comments are given in column[4].

\begin{figure}
\begin{center}
\plotone{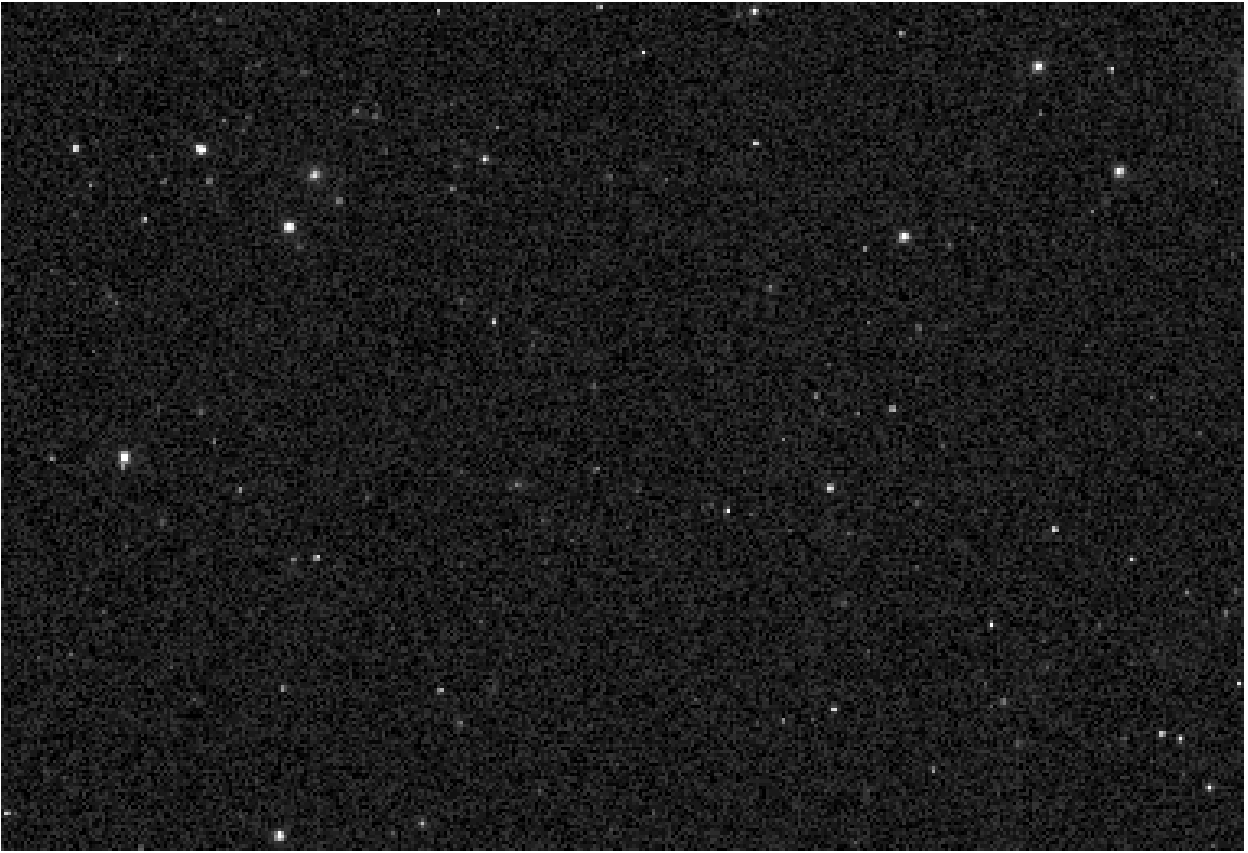}
 \caption{
 \label{fig:dwarf-region}
The successful example of Cut \& Enhance method.
The image is 6'$\times$13' true color image of the SDSS commissioning
 data.
There are many faint galaxies in the region.
Cut \& Enhance method has the ability to detect
the region in the sky where many faint galaxies are clustering.
This cluster was found only with Cut \& Enhance method.
}
\end{center}
\end{figure}
\begin{figure}
\begin{center}
\plotone{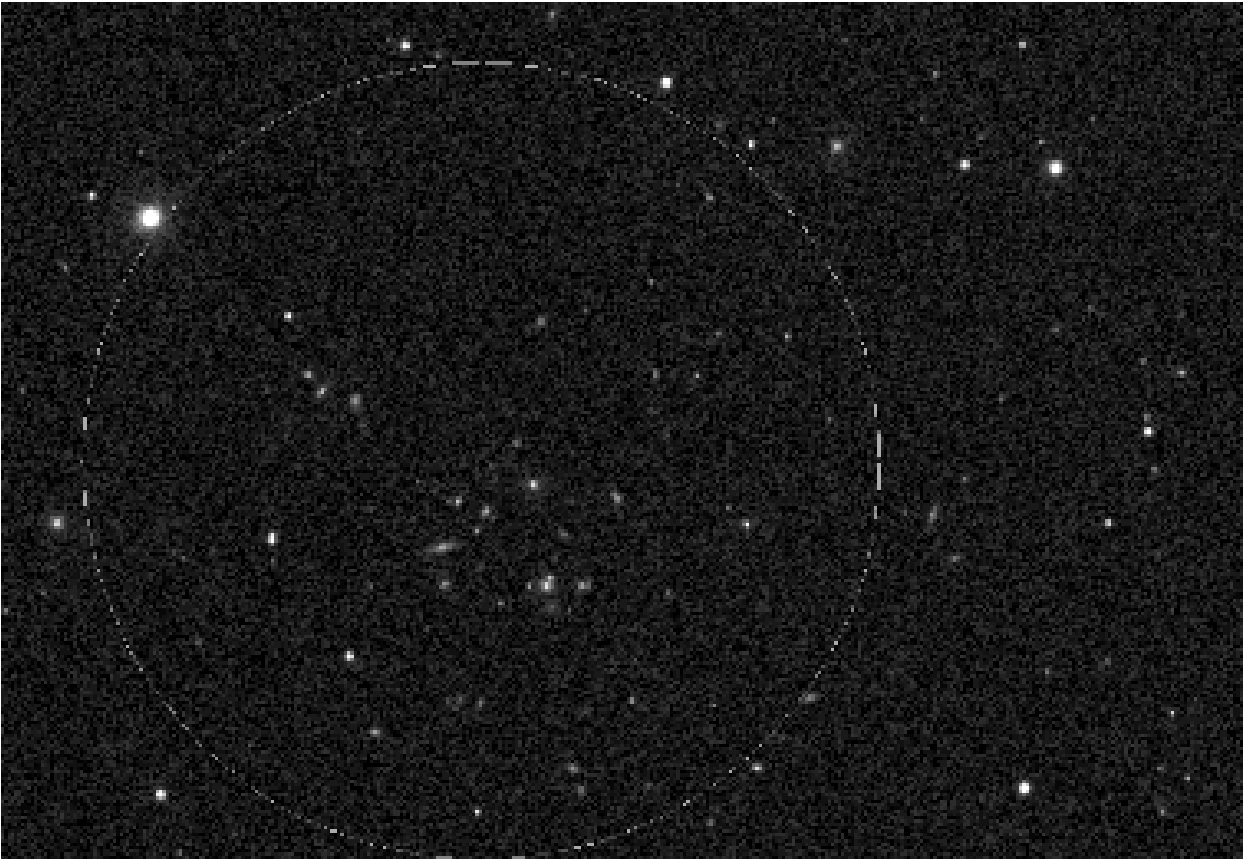}
 \caption{
 \label{fig:success-cluster}
The successful example of Cut \& Enhance method.
The image is 6'$\times$13' true color image of the SDSS commissioning data.
The cluster position and radius is shown with a yellow circle.
}
\end{center}
\end{figure}

 As the successful examples, 
 we show two typical examples of clusters detected only with the Cut \& Enhance method
but not with the other methods (discussed below).
 One is a clustering of blue galaxies. 
Since the Cut \& Enhance method does not reject blue
spiral galaxies, it can detect clustering of several blue spiral
galaxies. 
 Indeed, some of the detected clusters that we visually inspected 
 are clustering of blue galaxies.
The other is a clustering of numerous faint elliptical galaxies;
 in these regions faint elliptical galaxies 
 spread out over a large area ($\sim$ 0.01 deg$^2$) but with no bright cluster galaxies. 
Cut \& Enhance method detects these
regions successfully with a large radius. Fig.\ \ref{fig:dwarf-region}
shows the true color image of one of these clusters with numerous faint elliptical
galaxies.
 Fig.\ \ref{fig:success-cluster} shows a typical galaxy cluster successfully detected with Cut \& Enhance method.

%


%
%
%
%
%
%
%
%
%

\section{Comparison with other methods}\label{sec:compare}

 At the time of writing, the SDSS collaboration has implemented several
independent cluster finding methods and have run these algorithms on the SDSS
commissioning data. These methods include the Matched Filter (MF; Kim
et al.\ 2001), Voronoi Tessellation (VTT; Kim et al.\ 2001), and maxBCG
technique (Annis et al.\ 2001). Therefore, we have the unique opportunity to
compare the different catalogs these algorithms produce to further understand
each algorithm and possible differences between them. (also see Bahcall
et al. 2002 for comparisons of SDSS cluster catalogs.)

Here we provide a comparison between the CE method and the MF, VTT and maxBCG
techniques using a small sub--region of the SDSS data {\it i.e.} 23.75 deg$^2$
of commissioning data with RA between 16 and 25.5 degrees and Declination
between -1.25 and +1.25 degrees (The region in Fig. \ref{fig:beforecut_comparison}). We first matched the CE catalog with each of
the other three catalogs using a simple positional match criterion of less
than six arcminutes.  The number of matches between the CE and other catalogs
varies significantly because each cluster--finding algorithm has a different
selection function. At present, the selection functions for all these
algorithms are not fully established so we have not corrected for them in this
comparison.
 Although each algorithm measures cluster richness and redshift in its
 own way, the scatter between the measurement is large and it makes the
 comparison difficult.
 Therefore, we re-measured richness and redshift of the MF, VTT and maxBCG
 clusters using CE method
to see the richness and redshift dependence of the comparison.

\begin{figure}
\plotone{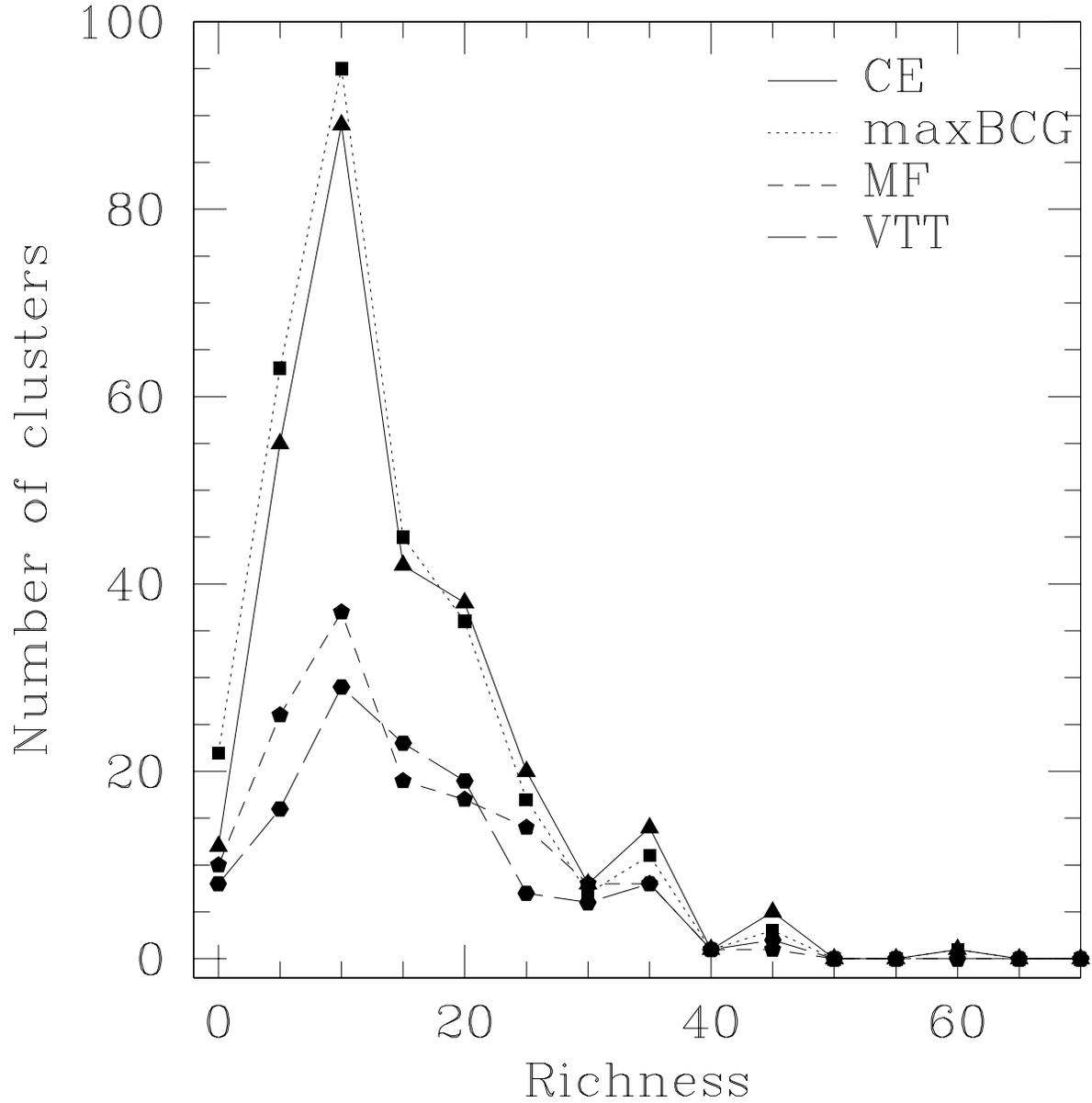}
\caption{
\label{fig:rich}
Comparison of four catalogs by richness. 
The abscissa is the richness of the cluster.
The ordinate is the number of the detected clusters.
Cut \& Enhance clusters are drawn with solid lines.
maxBCG clusters are drawn with dotted lines.
Matched Filter clusters are drawn with small dashed lines.
Voronoi tessellation clusters are drawn with long dashed lines.
Cut \& Enhance and maxBCG detect poor clusters (richness $<$20) more than MF or VTT.
}
\end{figure}

\begin{figure}
\plotone{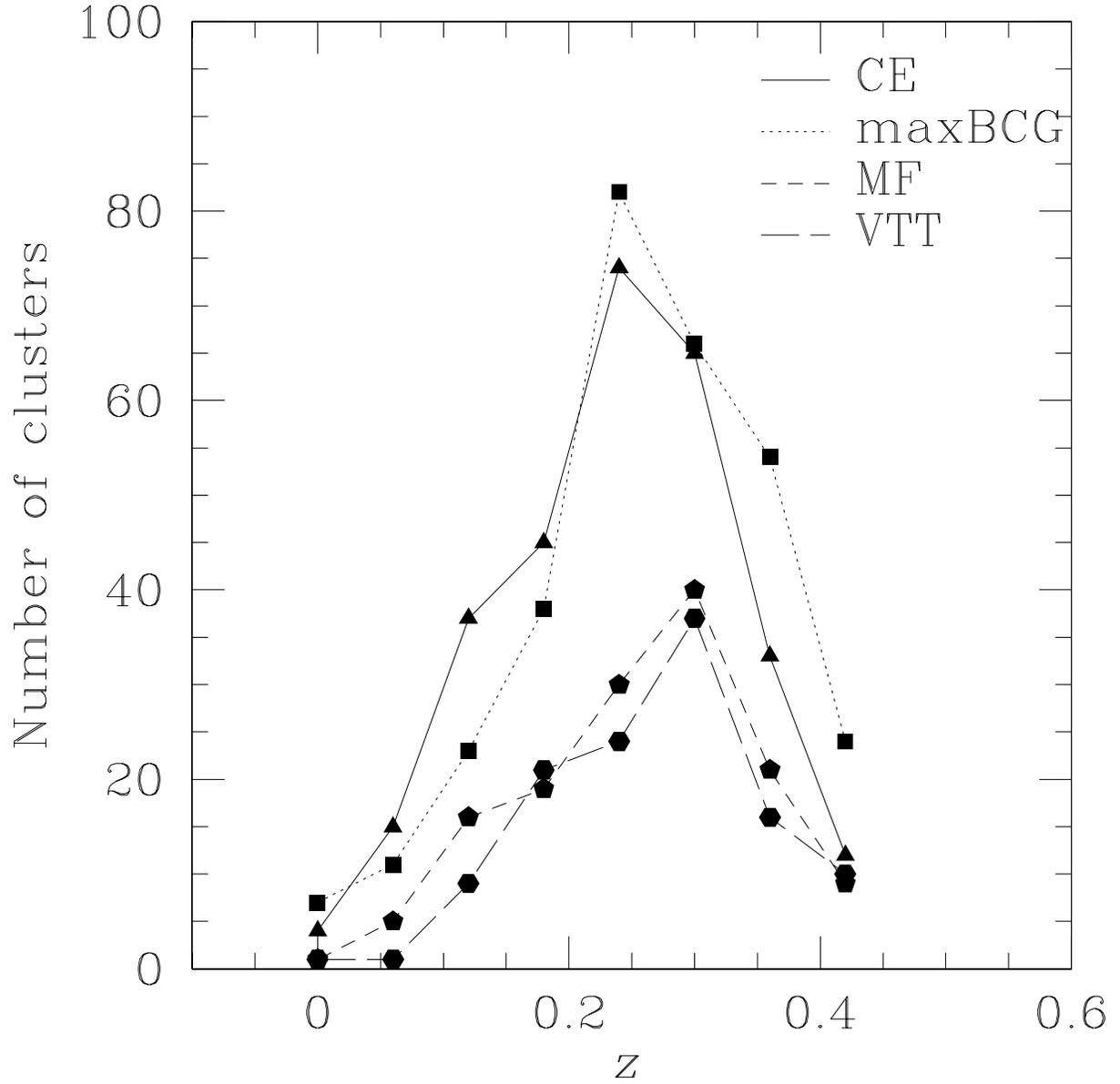}
\caption{
\label{fig:redshift}
Comparison of four catalogs by redshift.
The abscissa is the redshift of the clusters.
The ordinate is the number of the clusters.
Cut \& Enhance clusters are drawn with solid lines.
maxBCG clusters are drawn with dotted lines.
Matched Filter clusters are drawn with small dashed lines.
Voronoi tessellation clusters are drawn with long dashed lines.
The redshift is estimated using the color (described in Sect.2)
}
\end{figure}

In Table \ref{tab:number2}, we list the number of clusters each method finds
in our test region (Column 2 called ``N detection''). We also list in column 3
the number of the clusters found in common between the CE method discussed
herein and each of other method discussed above. For comparison, in Table
\ref{tab:other}, we also compare the number and percentage of matches found
between the VTT, MF and maxBCG technique.
 These two tables illustrate that the
overlap between all four algorithms is between 20 to 60\% which is simply a
product of their different selection functions. Furthermore, we note we have
used a simplistic matching criteria which does not account for the cluster
redshift or the errors on the cluster centroids. Future SDSS papers will deal
with these improvements (Bahcall et al. 2002). Tables \ref{tab:number2} \& \ref{tab:other} show that
the CE and maxBCG methods detect overall more clusters than the other methods
{\it i.e.}  363 and 438 clusters respectively, compared with 152 and 130
clusters for MF and VTT respectively.  This difference in the number of
clusters found is mainly due to differences in the thresholds used for each of
these algorithms. As illustrated in Fig.\ \ref{fig:rich}, a majority of the
extra clusters in the maxBCG and CE catalogs are lower richnesses
systems.
 As seen in Fig. \ref{fig:redshift}, these extra, lower richness,
systems appear to be distributed evenly over the entire redshift range of the
CE catalog ({\it i.e.} out to $z\simeq 0.4$).

\subsection{Comparison of Matched Filter and Cut \& Enhance Methods}\label{sec:mf-ce}

\begin{figure}\begin{center}	       
\plotone{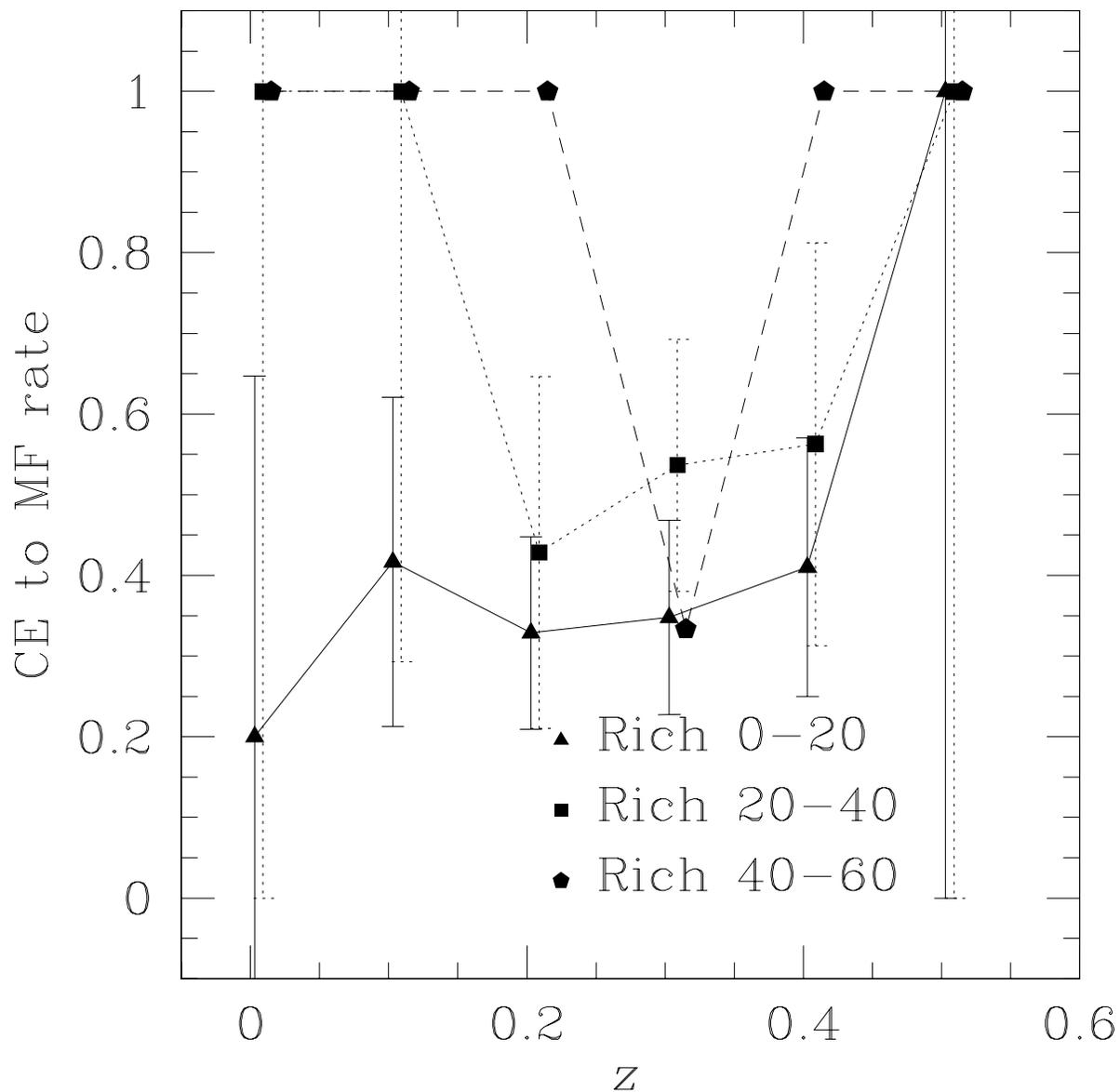}
	       \caption{
	       \label{fig: 0012030-tomomf-rate-z-rich.eps}
	       Comparison of MF with Cut \& Enhance.  The abscissa is the estimated
redshift. The ordinate is the rate of the MF clusters which are found
in Cut \& Enhance catalog to the number of the CE clusters.
 CE richness 0$\sim$20 is plotted with solid lines. CE richness 20$\sim$40
is plotted with dotted lines. CE richness 40$\sim$60 is plotted with dashed lines.
The error bars for richness 40$\sim$60 clusters are large and 
omitted for clarity (at $z$=0.3, the error is 80\%).  
The data for  richness 20$\sim$40 and 40$\sim$60 are shifted in redshift
	       direction by 0.01 for clarity.
}
	      \end{center}
\end{figure}

We focus here on the comparison between CE and the MF (see Kim et al. 2001).
In Fig.\ \ref{fig: 0012030-tomomf-rate-z-rich.eps}, we show the fraction of
MF clusters found in the CE catalog. We also split the
sample as a function of CE richness.  In Fig.\ \ref{fig:
0012030-mftomo-rate-z-rich.eps}, we show the reverse relationship {\it i.e.}
the fraction of CE clusters found by MF as a function of estimated redshift
and CE richness. These figures show that there is almost complete overlap
between the two catalogs for the highest redshift and richnesses systems in
both catalogs (there are however, only 5 $z>0.3$ systems in the MF catalog).
At low redshifts ($z<0.3$), the overlap decreases {\it e.g.} only 60\% of MF
clusters are found in CE catalog. To understand this comparison further, we
visually inspected all the clusters found by the CE method that were missing
for the MF catalog.  As expected, most of these systems were compact ($\sim$
1 arcminute) groups of galaxies.

\begin{figure}
\begin{center}
\plotone{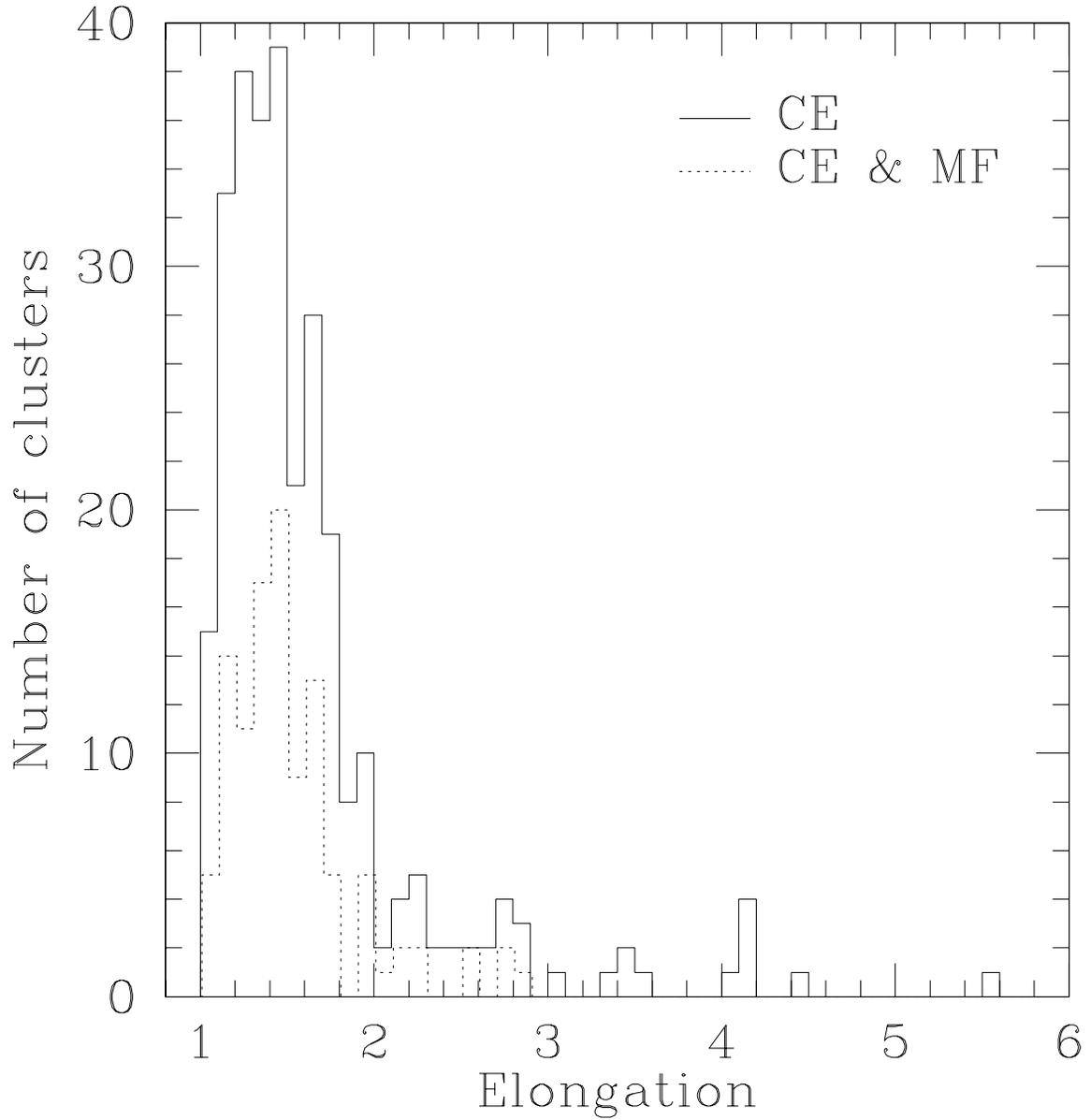}
\caption{
\label{fig:elong}
The elongation distribution of the detected clusters.
The Number of the clusters is plotted against
 the elongation of clusters (ratio of the major axis
 to minor axis).
The solid line is for the clusters detected with Cut \& Enhance method.
The dotted line is for the  the clusters detected with both Matched
 Filter and Cut \& Enhance method, which is shifted by 0.01 for clarity.
}
\end{center}
\end{figure}

Finally, in Fig.\ \ref{fig:elong}, we plot the distribution of axes ratios
(the major over the minor axis of the cluster) for both the whole CE catalog
as well as just the CE clusters found in MF catalog. This plot shows that a
majority of clusters in both samples have nearly spherical morphologies with
the two distributions in good agreement up to an axes ratio of 3 to 1.
However, there is a tail of 11 CE clusters which extends to higher axes ratios
that is not seen in the CE plus MF sub--sample. 
However, this is only $\sim$3\% of the CE clusters.
%

\subsection{Comparison of maxBCG and Cut \& Enhance Methods}

\begin{figure}[h]
\plotone{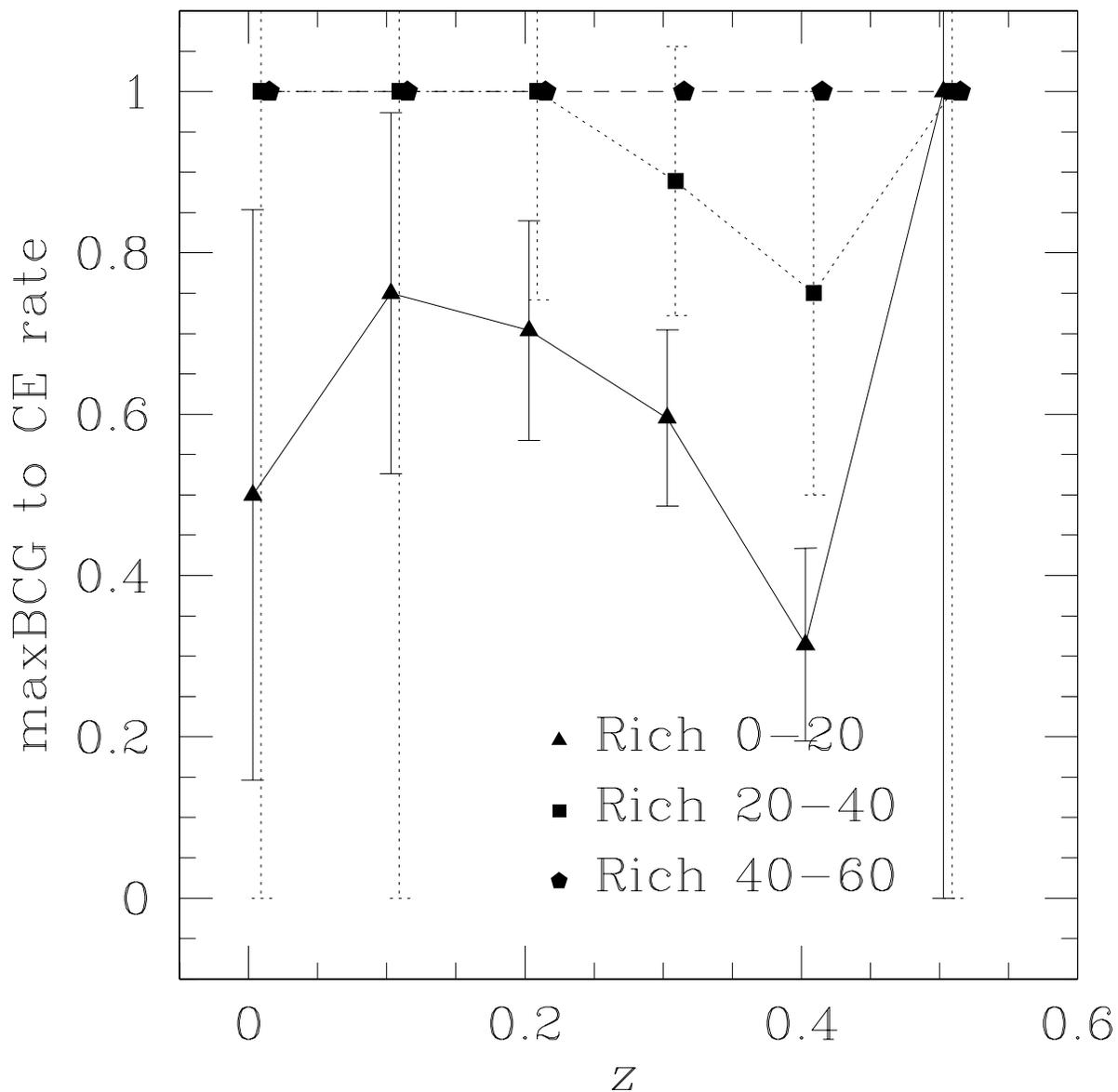}
\caption{
\label{fig:0012030-tomojim-rate-z-rich.eps}
Comparison of maxBCG clusters with Cut \& Enhance catalog.
 The abscissa is the color estimated
redshift. The ordinate is the ratio of the maxBCG clusters which are found
in Cut \& Enhance catalog to the number of the maxBCG clusters. The
 error bars for richness 40$\sim$60 clusters are large and omitted for
 clarity (at $z$=0.3, the error is 80\%).
The data for  richness 20$\sim$40 and 40$\sim$60 are shifted in redshift
	       direction by 0.01 for clarity.}
\end{figure}

In Fig.\ \ref{fig:0012030-tomojim-rate-z-rich.eps}, we show the fraction of
maxBCG clusters which are found in the CE catalog, while in Fig.\ \ref{fig:
0012030-jimtomo-rate-z-rich.eps}, we show the reverse relationship {\it i.e.}
the fraction of CE clusters found in the maxBCG catalog. In both figures, we
divide the sample by estimated redshift and observed CE richness.  First, we
note that the matching rate of maxBCG clusters to CE is $\sim70$\% or better
for clusters with a richness of $>20$ at all redshifts.  For the lower
richnesses systems, the matching rate decreases for all redshifts.  To further
understand the comparison between these two samples of clusters, we first
visually inspected all clusters detected by the CE method but were missing
from the maxBCG sample and found them to be blue, nearby poor clusters.  This
is a reflection of the wider color cuts employed by the CE method which allows
the CE algorithm to include bluer, star--forming galaxies into its color
criterion. The maxBCG however is tuned specifically to detect the E/S0
ridge--line of elliptical galaxies in clusters. We also visually inspected all
maxBCG clusters that were not found by the CE method and found these systems
to be mostly faint higher redshift clusters whose members mostly have fallen below the
magnitude limit used for CE method ($r^*$=21.5).

\subsection{Comparison of VTT and the CE Methods}

\begin{figure}[h]
\plotone{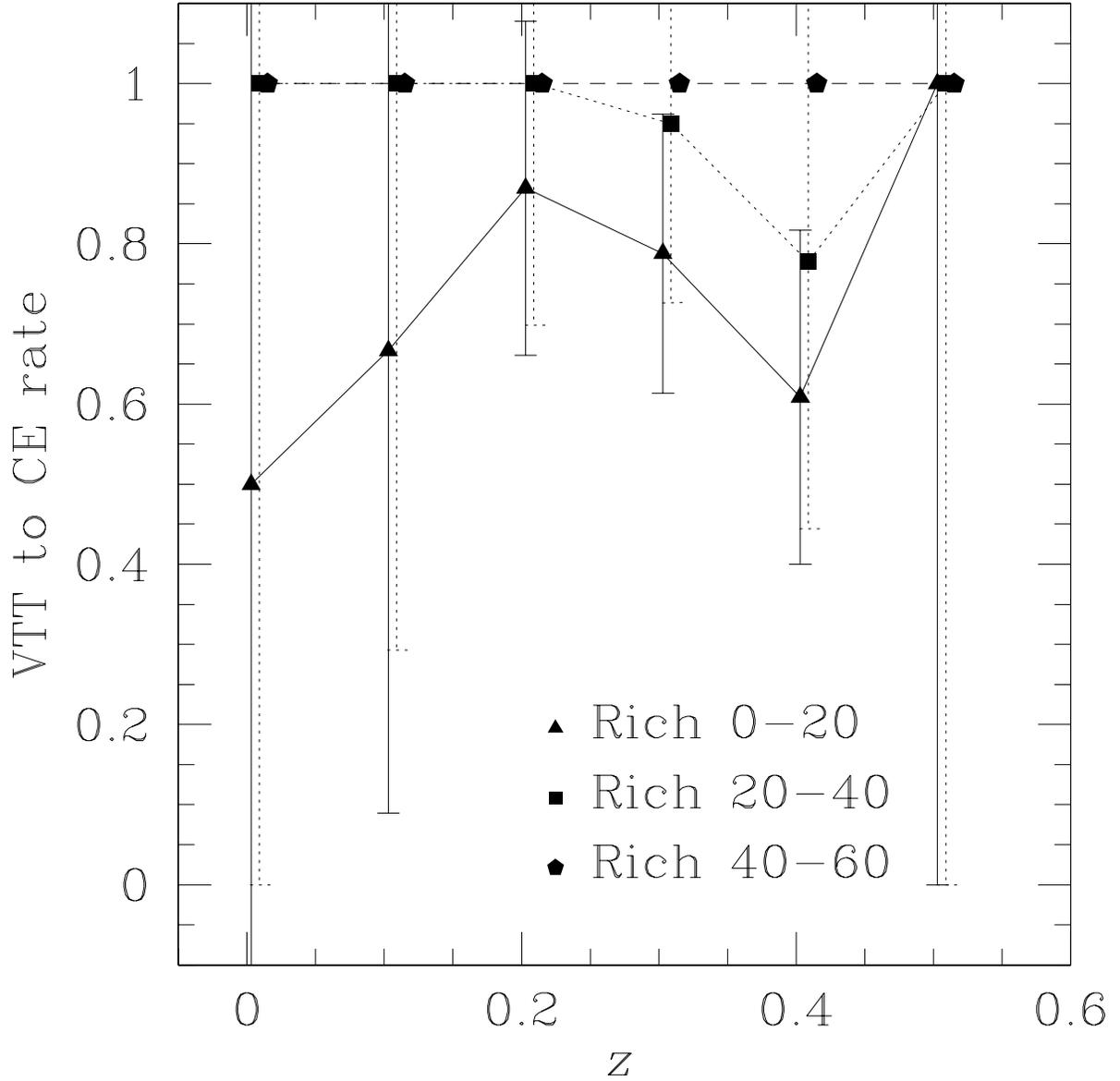}
\caption{
\label{fig: 0012030-vtttomo-rate-z-rich.eps}
Comparison of Cut \& Enhance with VTT. 
  The abscissa is the estimated
redshift. The ordinate is the rate of the Cut \& Enhance clusters which are found
in VTT catalog to the number of the Cut \& Enhance clusters. 
Note that Cut \& Enhance detects twice as many as VTT. The error bars
 for richness 40$\sim$60 clusters are 
large and omitted for clarity.
The data for  richness 20$\sim$40 and 40$\sim$60 are shifted in redshift
	       direction by 0.01 for clarity. }
\end{figure}

\begin{figure}
\plotone{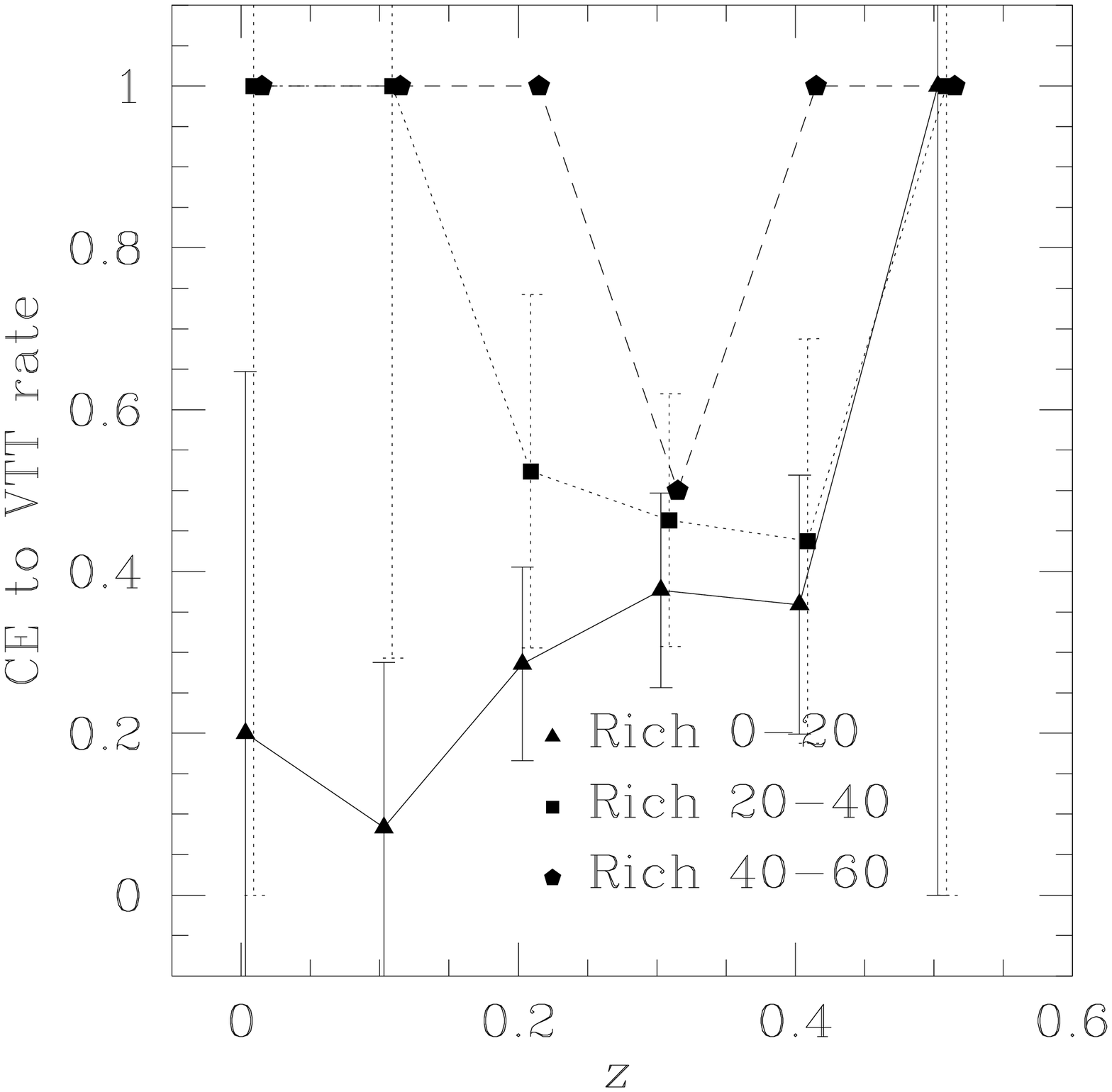}
\caption{
\label{fig: 0012030-tomovtt-rate-z-rich.eps}
Comparison of VTT with Cut \& Enhance. 
 The abscissa is the estimated
redshift. The ordinate is the ratio of the VTT clusters which are found
in Cut \& Enhance catalog to the number of the VTT clusters.
Note that Cut \& Enhance detects twice as many as VTT. The error bars for richness 40$\sim$60 clusters are large and omitted for clarity.
The data for  richness 20$\sim$40 and 40$\sim$60 are shifted in redshift
	       direction by 0.01 for clarity.}
\end{figure}

In Fig.\ \ref{fig: 0012030-vtttomo-rate-z-rich.eps}, we show the fraction of
 VTT clusters which were found by the CE as a function of estimated redshift
 and CE richness. Fig.\ \ref{fig: 0012030-tomovtt-rate-z-rich.eps} shows the
 fraction of CE found by VTT catalog as a function of estimated redshift and
 CE richness.  Because CE method detects twice as many clusters as does VTT,
 the matching rate is higher in Fig.\ \ref{fig:
 0012030-vtttomo-rate-z-rich.eps} than in Fig.\ \ref{fig:
 0012030-tomovtt-rate-z-rich.eps}, showing that the CE catalog contains a high
 fraction of VTT clusters.  In Fig.\ \ref{fig:
 0012030-tomovtt-rate-z-rich.eps}, the matching rate of low richness clusters
 improves at higher redshift because the poor clusters, which VTT does not
 detect become fainter and therefore both methods can not detect these
 clusters at high redshift.

\section{Summary} \label{sec:summary}

 We have developed a new cluster finding method, the Cut \& Enhance method.
It uses 30 color cuts and four color-color cuts
to enhance the contrast of galaxy clusters over the background galaxies.
 After applying the color and color-color cuts,
the method uses the color and angular separation weight of galaxy pairs
as an
enhancement method to increase the signal to noise ratio of  galaxy
clusters.
We use the Source Extractor to  detect galaxy clusters from the enhanced maps.
The enhancement and detection are performed for
every color cut, producing 34 cluster lists, which are then merged into
a single cluster catalog.

 Using the Monte Carlo simulations with real SDSS background
 as well as shuffled background,
 the Cut \& Enhance method is shown to
have  the ability to
detect rich clusters ($Ngal$=100) to $z\sim0.3$ with $\sim$80\%
 probability.
 The probability drops sharply at $z$=0.4 due to the flux limit of the
 SDSS imaging data.
The positional accuracy is better than 40'' for all richnesses examined
 at $z\leq$0.3. 
 The false positive test shows that over 70\% of clusters are likely to be real systems for CE richness $>$10. 
 We apply Cut \& Enhance method to the SDSS commissioning data and
 produce an SDSS Cut \& Enhance cluster catalog containing 4638 galaxy clusters in $\sim$350 deg$^2$. 
  We compare the CE clusters  with other cluster detection methods:
MF, maxBCG and  VTT. 
 The SDSS Cut \& Enhance cluster catalog developed in this work is a useful tool to study both cosmology and property of clusters and cluster galaxies.

\section{Acknowledgments}

We would like to express our sincere gratitude to 
Wataru Kawasaki, Kazuhiro Shimasaku,
Sadanori Okamura, Mamoru Doi, 
Chris Miller, Francisco Castander and  Shang-Shan Chon 
for their suggestions and discussions to improve the work.
We would like to thank anonymous referee whose comments significantly improved the paper. 
We are grateful to all the people who helped to build the Sloan Digital Sky Survey. 
T. G. acknowledges financial support from the Japan Society for the
Promotion of Science (JSPS) through JSPS Research Fellowships for Young Scientists.

The Sloan Digital Sky Survey (SDSS) is a joint project of The University of Chicago, Fermilab, the Institute for Advanced Study, the Japan Participation Group, The Johns Hopkins University, the Max-Planck-Institute for Astronomy (MPIA), the Max-Planck-Institute for Astrophysics (MPA), New Mexico State University, Princeton University, the United States Naval Observatory, and the University of Washington. Apache Point Observatory, site of the SDSS telescopes, is operated by the Astrophysical Research Consortium (ARC). 

Funding for the project has been provided by the Alfred P. Sloan Foundation, the SDSS member institutions, the National Aeronautics and Space Administration, the National Science Foundation, the U.S. Department of Energy, the Japanese Monbukagakusho, and the Max Planck Society. The SDSS Web site is http://www.sdss.org/.

\clearpage



\clearpage

%

\clearpage

\clearpage

%
%

\clearpage


%
%

%

%

%
%
%

%
%

\clearpage

%

%

\clearpage

%

\begin{figure}\begin{center}	       
\plotone{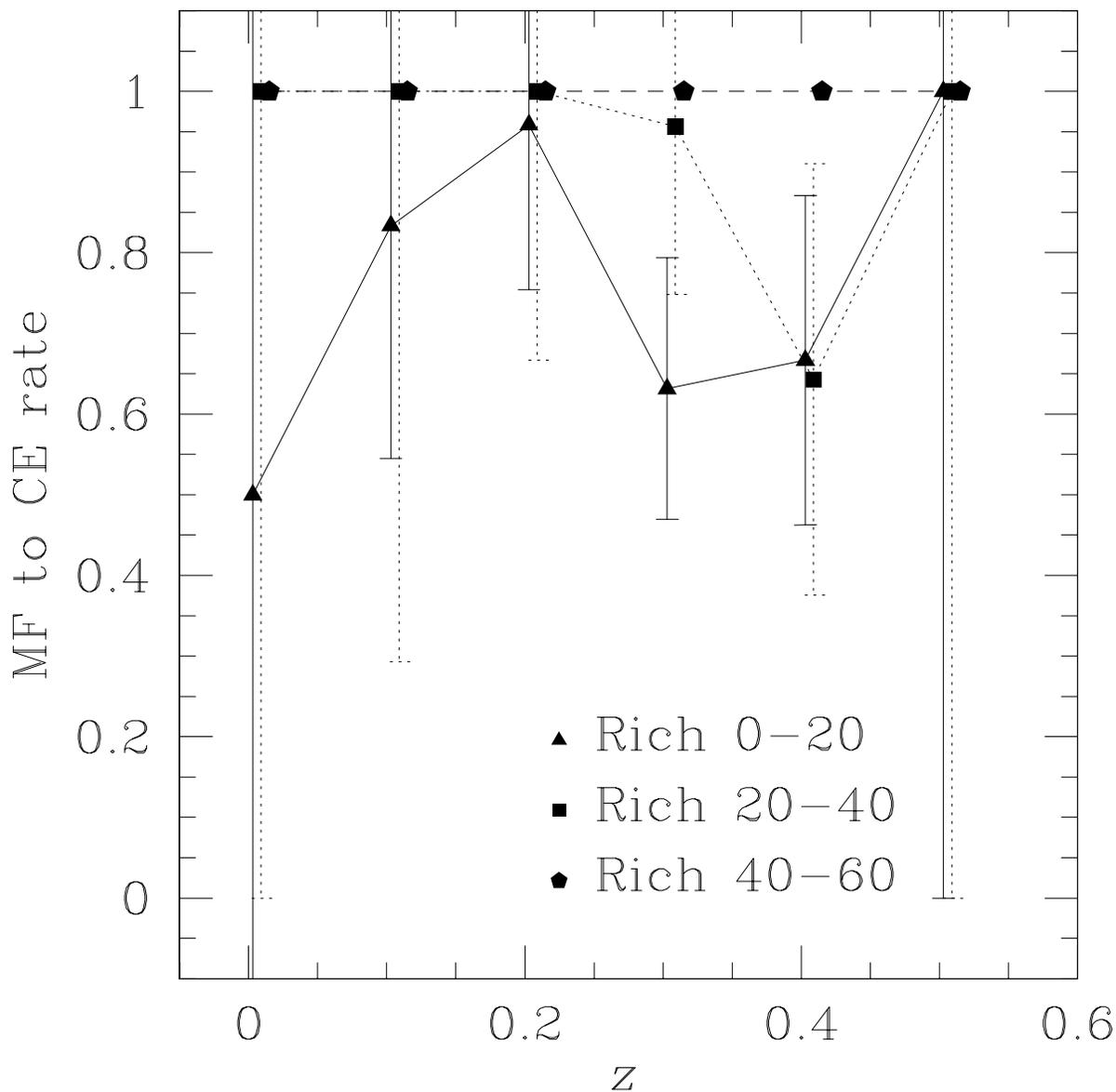}
	       \caption{
	       \label{fig: 0012030-mftomo-rate-z-rich.eps}
	       Comparison of Cut \& Enhance with MF. The abscissa is the estimated
redshift. The ordinate is the rate of the Cut \& Enhance clusters which are found
in MF catalog to the number of the Cut \& Enhance clusters. 
 Matching rate is low for poor clusters indicating
 	       Cut \& Enhance detects poor clusters more.
The error bars for richness 40$\sim$60 clusters are large and 
omitted for clarity (at $z$=0.3, the error is 80\%).
The data for  richness 20$\sim$40 and 40$\sim$60 are shifted in redshift
	       direction by 0.01 for clarity.
	      }\end{center}
\end{figure}

%
%

\clearpage

%
%

\begin{figure}[h]
\plotone{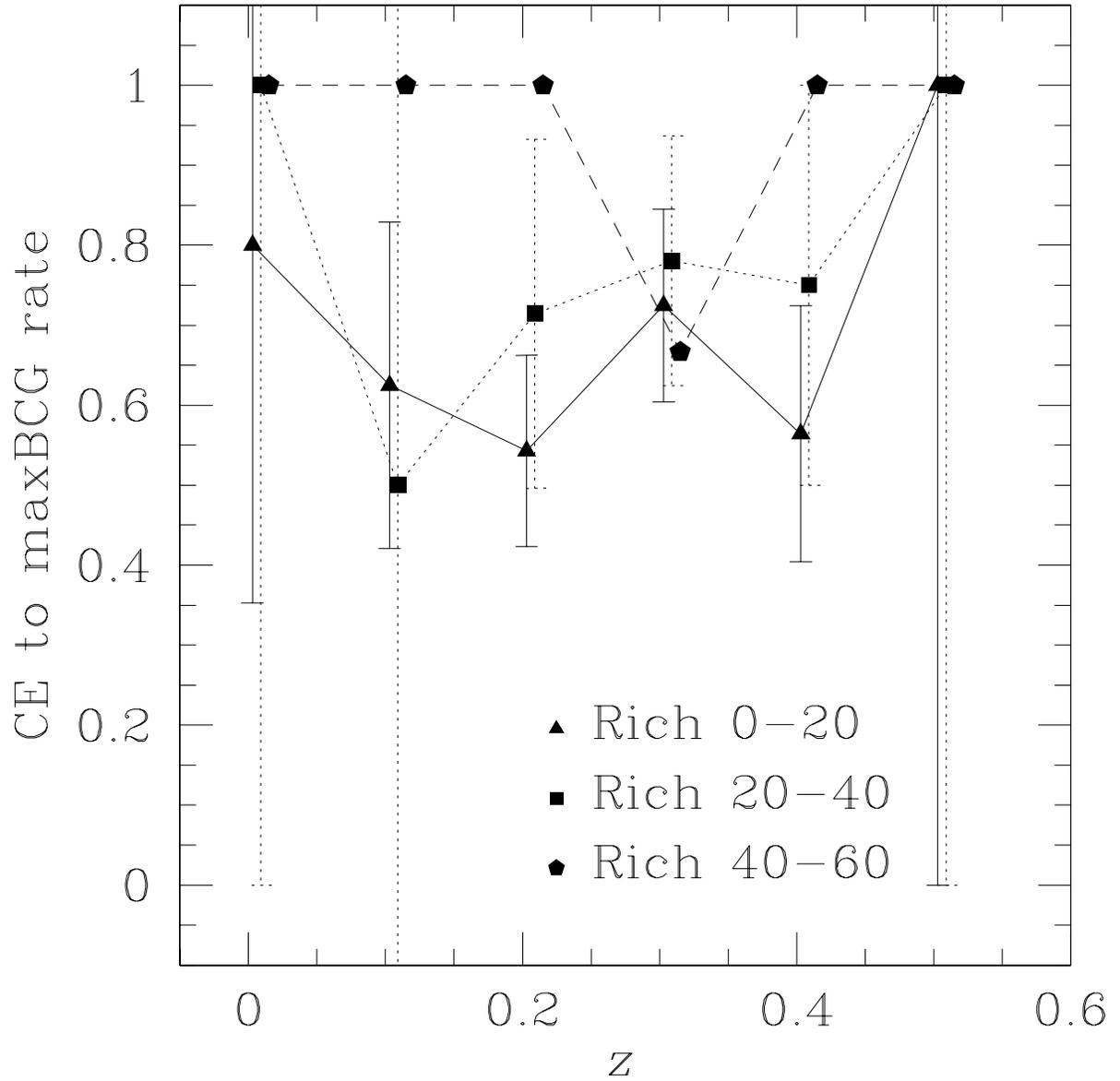}
\caption{\label{fig: 0012030-jimtomo-rate-z-rich.eps}
Comparison of Cut \& Enhance with maxBCG. The abscissa is the estimated
redshift. The ordinate is the rate of the Cut \& Enhance clusters which are found
in maxBCG catalog to the number of the Cut \& Enhance clusters. 
The error bars for richness 40$\sim$60 clusters are large and omitted for clarity.
The data for  richness 20$\sim$40 and 40$\sim$60 are shifted in redshift
	       direction by 0.01 for clarity.}
\end{figure}

%
%

\clearpage

%

%
%

\clearpage

%
%


%
%

\begin{table}[h]
\caption[ ]{\label{tab:cm-tilt}
Tilt of color--magnitude relation of A1577.
}
\begin{center}
\begin{tabular}{lllll}
\tableline
Color & Tilt (color/mag) & (magnitude range)  & Scatter (mag) &  (magnitude range) \\
\tableline
\tableline
$g^*-r^*$ &  0.0737 & $r^*<$19  & 0.081 & $r^*<$17   \\
$r^*-i^*$ &  0.0898 & $r^*<$19  & 0.040 & 18$<r^*<$19   \\
$i^*-z^*$ &  0.0018 & $r^*<$21  & 0.033 & 18$<r^*<$19   \\
\tableline
\end{tabular}
\end{center}
\end{table}

\clearpage

\begin{table}[h]
\caption[ ]{\label{tab:RXJ0256_color_cut}
The fraction of galaxies inside the color-cut inside of the
 RXJ0256.5+0006 and outside of RXJ0256.5+0006.
}
\begin{center}
\begin{tabular}{llll}
\tableline
Color cut & In cluster region(\%)  & Outside of cluster (\%) &   \\
\tableline
\tableline
$g^*-r^*$ &  36.9$^{+7.0}_{-6.0}$ & 13.57$\pm$0.03    \\
$r^*-i^*$ &  62.1$^{+8.8}_{-7.7}$ & 42.35$\pm$0.06    \\
$i^*-z^*$ &  59.2$^{+8.6}_{-7.6}$ & 44.55$\pm$0.06    \\
$g^*-r^*-i^*$ &  58.3$^{+8.5}_{-7.6}$ & 48.77$\pm$0.06   \\
$r^*-i^*-z^*$ &  76.7$^{+10.7}_{-7.7}$ & 65.68$\pm$0.07     \\
$g^*-r^*-i^*$ high$z$&  29.1$^{+6.3}_{-5.3}$ & 10.86$\pm$0.03    \\
$r^*-i^*-z^*$ high$z$&  6.8$^{+3.7}_{-2.5}$  & 9.94$\pm$0.02    \\
\tableline
\end{tabular}
\end{center}
\end{table}

\clearpage

\begin{table}[b]
\caption[ ]{\label{tab:sigma_test}
Sigma cut test. 
}
\begin{flushleft}
\begin{tabular}{l|lllll}
\tableline
Sigma    & 2 &  4 & 6 & 8 & 10 \\
\tableline
N detection &  402 &   437   &  453 &  434 & 415   \\
\tableline
\end{tabular}
\end{flushleft}
\end{table}

\clearpage

\begin{table}[b]
\caption[ ]{\label{tab:fluxmax}
Test of fluxmax=1000 cut. 
}
\begin{flushleft}
\begin{tabular}{l|lllll}
\tableline
Fluxmax    & 500 &  0750 & 1000 & 1500 & 2000 \\
\tableline
N detection &  890 &   655   &  464 & 260 &10   \\
\tableline
\end{tabular}
\end{flushleft}
\end{table}

\clearpage

\begin{table}[h]
\caption[ ]{\label{tab:ce-miss-new}
10 false detections of Cut \& Enhance method.
The region used is RA between 16 and 25.5 deg, DEC between -1.25 and $+$1.25 deg,
 (23.75 deg$^2$).
 $\sigma$ (column [1]) is the significance of the detection. 
CE richness (column [2])  is the richness of the detection.
$z$ (column [3]) is the color estimated redshift of the detection.
Comment (column [4]) is the comment on the detection.
}
\begin{center}
\begin{tabular}{llll}
\tableline
$\sigma$ & richness  & $z$ & comment \\
\tableline
\tableline
12.39 &  31 & 0.22 & looks like field.\\
7.85 & 21 & 0.18 &  looks like field.\\
4.80 & 11 & 0.10 & looks like field.\\
16.74 & 16 & 0.18 & looks like field.\\
56.85 & 8 & 0.04 & a big galaxy.\\
9.35 & 7& 0.00 &   looks like field.\\
7.06 & 1& 0.04 &  eight blue galaxies.\\
11.0 & 14 & 0.04 &  looks like field.\\
 4848.39 & 17 & 0.12 & a big galaxy.\\
25.20 & 7 & 0.00 & a big galaxy.\\
\tableline
\end{tabular}
\end{center}

\end{table}

\clearpage

\begin{table}[b]
\caption[ ]{\label{tab:number2}
Ratio of number of clusters detected with MF, maxBCG and VTT to Cut \&
 Enhance clusters.
The region used is RA between 16 and 25.5 deg, DEC between -1.25 and $+$1.25 deg,
 (23.75 deg$^2$).
Column 1 lists the method. 
N detection (column [2]) is the numbers of clusters detected by each
 method.
Common detection (column  [3]) is the number of clusters detected by both
 the method and Cut \& Enhance method.
Rate to CE (column  [4]) is the percentile of the numbers of detection with each
 method to the numbers of detection with Cut \& Enhance method (CE in
 the table).
Rate to the method (column [5]) is the percentile of the numbers of detection with
 Cut \& Enhance method to the numbers of detection with each method.
}
\begin{flushleft}
\begin{tabular}{llllll}
\tableline
\tableline
 & N detection  & Common detection &
 Rate to CE (\%)  & Rate to the method (\%) \\
\tableline
MF & 152 & 116  &   32.0 & 76.3  \\
\tableline
maxBCG & 438 & 183 & 50.4 & 41.8\\
\tableline
VTT & 130 & 96 & 26.4 &73.8\\
\tableline
\tableline CE & 363 & - & - &- \\
\tableline
\end{tabular}
\end{flushleft}
\end{table}

\clearpage

\begin{table}[b]
\caption[ ]{\label{tab:other}
The comparison of the detected clusters by other methods than Cut \&
 Enhance method.
Column 1 and row 1 denote the names of each method.
Total numbers of the clusters detected with each method in the region RA
 between 16 and 25.5 deg, DEC between $-$1.25 and  $+$1.25 deg, (23.75
 deg$^2$) are written in the parenthesis in row 1.
Rows  2$\sim$4 list the numbers of the clusters detected with both methods
 (column 1 and row 1) and their percentile to the methods in column 1. 
}
\begin{flushleft}
\begin{tabular}{l|lll}

\tableline
 & MF &  VTT & maxBCG \\
 & (152) &  (130) & (438) \\
\tableline
MF &  - &   39.4\% (60)   & 59.2\% (90)   \\
\tableline
VTT &  45.5\% (60) & - & 65.4\% (85) \\
\tableline
maxBCG &  20.5\% (90) & 19.4\% (85)&-\\
\tableline
\end{tabular}
\end{flushleft}
\end{table}

\clearpage

%
%

\end{document}